\documentclass[letterpaper, 11pt]{article}

\usepackage[english]{babel}

\usepackage[a4paper,top=1in,bottom=1in,left=1in,right=1in,marginparwidth=1in]{geometry}

\usepackage{amsmath}
\usepackage{graphicx}
\usepackage{setspace}
\usepackage{multirow}
\usepackage{algorithm}
\usepackage[noend]{algpseudocode}
\usepackage{csquotes}
\usepackage{subcaption}
\usepackage[noblocks]{authblk}

\usepackage{tabularx}
\usepackage{booktabs}
\usepackage{breakcites}

\providecommand{\keywords}[1]
{\small\textit{Keywords:}#1}

\usepackage{placeins} 
\newcolumntype{Y}{>{\raggedleft\arraybackslash}X}

\usepackage{xcolor}

\usepackage[font=small,labelfont=bf]{caption}

\newcounter{noteMCctr} 

\newcounter{noteGRctr} 

\newcommand{\gdr}[1]{{\color{black} #1}}
\newcommand{\gr}[1]{{\color{black} #1}}

\definecolor{colour3}{RGB}{178,55,250} 
\newcommand{\mc}[1]{{\color{black} #1}}

\newcommand{\mcc}[1]{{\color{black} #1}}

\newcommand{\update}[1]{{\color{black} #1}}

\usepackage{booktabs}
\newcommand{\bftab}{\fontseries{b}\selectfont}

\usepackage[colorlinks=true, allcolors=blue, breaklinks=true]{hyperref}
\usepackage{cleveref}

\usepackage{amsfonts}
\usepackage[citestyle=bwl-FU, uniquename=true]{biblatex}

\graphicspath{ {./graphs/} }

\usepackage{tcolorbox}

\title{
Co-trading networks for modeling dynamic interdependency structures and estimating high-dimensional covariances in US equity markets}

\author[1]{Yutong Lu}
\author[1,3]{Gesine Reinert}
\author[1,2,3,4]{Mihai Cucuringu}

\affil[1]{\small Department of Statistics, University of Oxford, Oxford, UK}
\affil[2]{\small Mathematical Institute, University of Oxford, Oxford, UK}
\affil[3]{\small The Alan Turing Institute, London, UK}
\affil[4]{\small Oxford-Man Institute of Quantitative Finance, University of Oxford, Oxford, UK}

\begin{document}

\maketitle

\begin{abstract}
\noindent The time proximity of trades across stocks reveals interesting topological structures of the equity market in the United States. In this article, we investigate how such concurrent cross-stock trading behaviors, which we denote as \textit{co-trading}, shape the market structures and affect stock price co-movements. By leveraging a co-trading-based pairwise similarity measure, we propose a novel method to construct dynamic networks of stocks. Our empirical studies employ high-frequency limit order book data from 2017-01-03 to 2019-12-09. By applying spectral clustering on co-trading networks, we uncover economically meaningful clusters of stocks. Beyond the static Global Industry Classification Standard (GICS) sectors, our data-driven clusters capture the time evolution of the dependency among stocks. Furthermore, we demonstrate statistically significant positive relations between low-latency co-trading and return covariance. With the aid of co-trading networks, we develop a robust estimator for high-dimensional covariance matrices, which yields superior economic value on portfolio allocation. The mean-variance portfolios based on our covariance estimates achieve both lower volatility and higher Sharpe ratios than standard benchmarks. \\
    
\noindent\keywords{ Market microstructure; Co-occurrence analysis; Network analysis; Machine learning; Robust covariance estimation; Portfolio allocation}
\end{abstract}


\section{Introduction}
The pioneering work of \cite{kyle1985continuous} posits that the price formation at high-frequency level is the result of the interplay among market participants. In this model, market makers monitor the aggregated order flows after informed and liquidity traders submit their orders, and then set their fair prices. With the development and boom of high-frequency trading (HFT) strategies, the interplay becomes more aggressive 
and sophisticated. Previous works 
(\cite{ brunnermeier2005predatory, van2019high, hirschey2021high, yang2020back, goldstein2023high}) find that \gr{some} HFT traders actively detect activities of other participants in the market, and fiercely trade against them. Furthermore, these interactions can span across different stocks (\cite{hasbrouck2001common, bernhardt2008cross, capponi2020multi}). We investigate concurrent (almost instantaneous) trading across multiple stocks, a phenomenon which we refer to as \textit{co-trading} behavior, at a very granular level by directly considering individual trades, and zooming-in around their local neighborhoods.

In this paper, we propose a novel method that constructs  \textit{co-trading networks}, in order to model the complex structures of co-trading activities in equity markets. Constructed from limit order books, our co-trading networks bridge the trading behaviors at a granular level with the dynamic topological structures of the equity market and price co-movements among individual stocks. Moreover, by making use of co-trading networks, we develop a robust estimator for high-dimensional covariance matrices, and demonstrate  its conspicuous economic value on portfolio allocation. 
 
The network construction starts with a pairwise similarity measure between stocks. Inspired by the idea of trade co-occurrence originating from \cite{lu2022trade}, we define the pairwise similarity as the normalized count of times that trades, for a given pair of stocks, arrive concurrently. We name this measure as a \textit{co-trading score}, since it embeds the intuition that stocks frequently traded together are closely related. Concatenating co-trading scores between every pair of stocks, we obtain the co-trading matrix, which serves as the 
\gr{adjacency matrix} of the proposed \gr{weighted} network of equity markets. 

Employing 
algorithms and 
tools for network analysis 
(\cite{hagberg2008exploring}), we provide empirical evidence that co-trading networks are capable of capturing meaningful structures of the equity market. By visualizing 
a co-trading network, aggregated over the entire period of study, with information filtering (\cite{mantegna1999hierarchical}), we observe that stocks from the same sector groups tend to have strong co-trading relations. \gr{This observation} 
echos 
\gr{the phenomenon that often} stocks in the same sector groups \gr{of the market} are likely to appear in the same portfolios and their prices tend to move together. Additionally, we uncover clusters, \gr{or communities,} of stocks in the co-trading network. 
To detect these communities, we apply a spectral clustering algorithm on the co-trading matrix to group stocks with similar co-trading behaviors into clusters. For comparison,
we select the Global Industry Classification Standard (GICS) as sector labels. Our empirical results show substantial overlap between the data-driven clusters and GICS sectors, which confirms that the co-trading network accommodate the sector structures as expected.
Moreover, we leverage the uncovered network to study the influence of both individual stocks and sectors on the market. By using tools such as  eigenvector centrality (\cite{bonacich1972factoring}, \cite{bonacich1987power}), we identify that large technology companies and financial institutions, such as Microsoft, Apple, \gr{and} JPMorgan, 
present 
\gdr{strong} co-trading relations with others, and thus have 
\gdr{high} impact on the structure of the market. Despite the alignment with GICS sectors, the co-trading network 
contains information beyond sectorial structure. The data-driven clusters also group together closely related stocks from different sectors.   

Apart from the aforementioned static graph, the construction of co-trading matrices is flexible \gr{due to the choice of distance in time at which two trades are deemed to co-occur.}
It is \gr{also} informative to analyze the time evaluation of co-trading networks and explore the dynamic of market structures. In this study, we focus on network time series at daily level, 
\gr{but this approach} can easily \gr{be} generalized to intraday, monthly and so forth. Our empirical findings from daily co-trading networks indicates that the co-trading relations across different sectors increase from 2017 to 2019. Using spectral clustering, we detect clusters at a daily level and compare them with GICS sectors. By \gdr{A plot of} 
the similarity across time \gdr{reveals}
a downward trend with fluctuations. In addition, we also compare the similarity among daily clusters. The variation in clusters also increases as the spread of co-trading beyond sector groups. By applying 
\gdr{a} spectral clustering algorithm (\cite{shi2000normalized, ng2002spectral, cucuringu2019sponge, cucuringu2020hermitian}) for change-point detection based on the temporal similarity heatmap, we uncover three distinct regimes over the period of study.

To exploit the association between co-trading behaviors and price co-movements, we conduct network regression analysis. We build realized covariance matrices from intraday returns as proxies for co-movements among stocks. By regressing covariance matrices against co-trading networks on a daily basis, we reveal  positive associations between the two types of matrices. On $98.51 \%$ of the days in the study, the positive regression coefficients are statistically significant at $5 \%$ significance level. 
Further controlling for GICS sectors, we conduct multiple regressions by adding a fixed sector network as an independent variable. Even with the presence of sectors, the positive and significant relation still holds. \update{Furthermore, after removing price co-movements \mcc{that} arise from factor structures, the co-trading matrices capture patterns in \mcc{the} idiosyncratic covariance among stocks}. Therefore, \gr{there is a strong indication that} the co-trading behaviors at the high-frequency level are positively correlated with the return covariance, and have 
\gdr{explanatory} power on price co-movements that goes beyond the \update{common risk factors} and commonly adopted sectors. 

With the aid of co-trading matrices, we propose a robust estimator for the high-dimensional covariance matrix of stock returns at daily frequency. By assuming that  stock returns follow a linear factor structure (\cite{ross1976arbitrage}), we  decompose the  covariance matrix as the sum of \gr{the} factor covariance and \gr{the}  idiosyncratic covariance. Then, we impose a block structure on the diagonal of the idiosyncratic covariance matrix, such that elements outside the blocks are set to zero. Our approach extends the work of \cite{ait2017using}, which uses principal component analysis to derive latent factors and form blocks using GICS sectors. However, \gr{as the GICS sectors are fixed,} the sector blocks remain static over time. As we have shown, the market structures are time-varying, and thus static sector memberships  \gdr{do not suffice} 
to capture the similarity of stocks at higher frequency. Therefore, we update the diagonal block structure daily with data-driven clusters derived from co-trading networks. To evaluate the performance, we construct mean-variance portfolios, subject to various leverage constraints. Our covariance estimators, based on dynamic clusters, generally outperform the baselines, using fixed GICS sectors, by achieving lower volatility and higher Sharpe ratios. Our best-performing portfolio achieves a Sharpe ratio of 1.40, which is 0.43 higher than the corresponding baseline \gdr{of 0.97, thus giving a 44\% improvement}, and 0.57 higher than the market \gdr{(0.83)} over the same period.  

The remainder of this paper is organized as follows. \Cref{sec:literature_review} outlines 
\gdr{the} contributions \gdr{of this work} to the existing literature. \Cref{sec:co-trading_network} introduces the definition of co-trading score and the construction of co-trading networks. In \Cref{sec:network_analysis}, we begin our empirical studies with conducting exploratory analysis on a static co-trading network and detecting clusters. Next, we study the dynamics of the daily co-trading matrices in \Cref{sec:temporal_networks}. Subsequently, we explore the relation between daily co-trading networks and covariance matrices in \Cref{sec:co-trading_covariance} and propose a co-trading based covariance estimator in \Cref{sec:covariance_estimation}. Then, \Cref{sec:robustness} provides a robustness analysis of \gr{the} proposed estimator.  
Finally, in \Cref{sec:conclusion}, we conclude and discuss limitations and potential research directions. \update{The Appendices contain details of the spectral clustering algorithm and robustness analysis.}

\section{Literature review} \label{sec:literature_review}
This study sits at the confluence of three strands of literature. Firstly, our research enriches the network analysis and the modeling of complex inter-dependency relations in financial markets. Network analysis has been proven to be effective in studying inter-dependency relations in complex systems\gdr{; in particular,}
there is a large literature on networks in financial markets (\cite{bardoscia2021physics, marti2021review}). 

Previous research develops various methodology to build financial networks. 
\gdr{T}he influential work of \cite{mantegna1999hierarchical} first builds a network from a distance measure based on correlations of stock returns, \gdr{which is then} filtered with a minimum spanning tree (MST). 
\gdr{M}any works followed by constructing networks with diverse distance \gdr{and} similarity measures, such as return correlations (\cite{tumminello2007correlation, di2010use}), Granger causality (\cite{billio2012econometric}), mutual information (\cite{fiedor2014information}), co-jumps of stock prices (\cite{ding2021stock}), and so forth. Additionally, multiple methods have been used to replace MST for information filtering, including random matrix theory (\cite{plerou2000random}), \mc{Potts} super-paramagnetic transitions (\cite{kullmann2000identification}), planar maximally filtered graph (\cite{tumminello2005tool, massara2016network}), \gdr{a} threshold-filter method (\cite{huang2009network, namaki2011network}), \gdr{and more}. 
Recent works apply deep learning and natural language processing techniques to construct networks of various types of relations among stocks, such as company competition, product similarity, \gdr{and} news co-occurrence, 
from diverse data sources, such as financial reports, media news, \gdr{and} events 
(\cite{hoberg2016text, schwenkler2019network, cheng2022financial, zhang2023company}).

Regarding networks of financial markets, preceding studies uncover the topological structures of markets through community detection. For example, the pioneering work of \cite{mantegna1999hierarchical} applies hierarchical clustering and discovers hierarchical structure stock portfolios. Moreover, the networks can be constructed as a time series (\cite{mcdonald2005detecting, nie2017dynamics}). 
\cite{bennett2022lead} and \cite{zhang2023robust} build lead-lag networks with different similarity measures, test multiple clustering algorithms and study 
time-varying lead-lag structures in the market.

To this field, our contribution is \gdr{firstly, the proposal of} 
an original similarity measure, directly \mc{derived from very granular} records of trades, with explicit interpretation as how frequently two stocks are traded together, \mc{with the final goal of constructing networks}. In addition, we detect dynamic clusters and provide a comprehensive comparison with GICS sectors.    

Secondly, this research contributes to the studies of market microstructure, especially interplay among trading activities. 
\cite{kyle1985continuous} posits a famous two-period model of high-frequency price formation by solving the equilibrium of the game between liquidity takers and market makers. 
Furthermore, various studies show that high-frequencies strategies can be aggressive. \cite{hirschey2021high} claims that high-frequency traders (HFTs) can predict order flows from other market participants and trade ahead of them. \cite{van2019high} provide empirical evidence that HFTs can detect the trading activities of institutional traders, and adjust their own \gdr{trading} strategies. 
Moreover, HFTs even actively explore the market by initiating small trades and watch\gdr{ing} the response of others (\cite{clark2013exploratory}). Researchers also propose theoretical models for the interactions between HFTs and other traders (\cite{grossman1988liquidity, brunnermeier2005predatory, yang2020back}). 

The interactions can span across different stocks on the market. \cite{bernhardt2008cross} state that the strategical interplay among speculators is often concurrent and across many stocks. There is vast literature on cross-impact (\cite{pasquariello2015strategic, benzaquen2017dissecting, schneider2019cross}), showing order flow of a stocks can affect prices of other stocks. 
\cite{lu2022trade} classify trades of stocks by whether they concurrently arrive with other trades for the same or different, or both same and different stocks, and investigate price impact using order imbalance from different groups of trades. They discover that the time proximity of trade arrivals
\gdr{has explanatory power for}
stock returns.

This study 
\gdr{uses the paper} \cite{lu2022trade} \gdr{as starting point but instead considers} 
interactions between the trades of every pair of stocks. Rather than studying the price impact of order imbalance on one stock at a time, we construct co-trading networks of all stocks 
\gdr{jointly} and 
\gdr{study} the impact of trading interactions at market microstructure level on macroscopic price co-movements.

Finally, our study adds to the \mc{growing body of financial econometrics literature} on robust estimation of high-dimensional covariance matrices of stock returns. An invertible and well-behaved covariance matrix is essential for portfolio allocation with mean-variance optimization (\cite{10.2307/2975974}). However, when the number of sampled timestamps is small relative to the size of the panel of stocks, sample covariance matrices \update{can be} singular or ill-conditioned. To overcome this issue, previous studies develop different streams of regularized estimation methods, including thresholding (\cite{bickel2008regularized, bickel2008covariance}), shrinkage (\cite{ledoit2003improved, ledoit2004well,chen2010shrinkage}), etc. 
%
These regularization techniques are based on structural assumptions; for example, thresholding estimators assume the covariance matrices are sparse. In particular for stocks, a large literature of asset pricing studies reveal linear factor structures of equity returns (\cite{sharpe1964capital, ross1976arbitrage, fama1992cross, fama1993common, fama2015five}). Based on factor models, the covariance matrix of returns can be decomposed as sum of a low-rank factor component and a residual idiosyncratic component. \mc{Various lines of work have then imposed} different types of sparsity assumptions \mc{on} the residual component which represents the covariance of idiosyncratic risk of stocks. \cite{fan2016overview} provide a through overview of factor-based robust covariance estimation. 
\mc{By using observable and latent factors, respectively, the works of
\cite{fan2016incorporating} and \cite{ait2017using} impose block structures on the idiosyncratic component, where stocks are sorted by their GICS sector membership}, thus forcing the residual covariance of stocks in different sectors to be zero. Consequently, they conclude that incorporating clusters with economic interpretation benefits the covariance estimation task. 

Our method for robust estimation \mc{contributes to this body of literature, and can be construed as a direct extension of the two aforementioned articles, by taking into account very granular high-frequency data that encodes higher-order relationships on the co-trading behavior. Instead of employing static GICS sectors, we perform time-varying data-driven clustering. By capturing the dynamic dependency relations between the universe of stocks, our proposed method outperforms baselines in portfolio allocation tasks}.

\section{Construction of co-trading networks} \label{sec:co-trading_network}
In this section, we propose pairwise co-trading scores to measure the similarity between two stocks, using the  \textit{co-occurrence of trades} methodology from \cite{lu2022trade}. Then we leverage these scores to build affinity matrices and construct co-trading networks.   

\subsection{Co-occurrence of trades}
We start with introducing notations and then define co-occurrence of trades. Here, we denote each trade as a 4-tuple. Let $x_k = (\tau_k, s_k, d_k, q_k)$ denote the \gr{information of the} $k^{th}$ trade, \gr{where} we capture the following trade characteristics
\begin{itemize}
    \item $\tau_k$ is the time when the trade occurs;
    \item $s_k$ indicates 
    \gr{name} of the stock \gr{for which the trade is} 
    executed; 
    \item $d_k \in \{ buy, sell \}$ indicates whether the trade is buyer-- or seller--initiated;
    \item $q_k$ is the volume of the trade.
    \end{itemize}
Then, for every trade $x_k$, we 
define a $\delta$-neighbourhood \gr{of trades which are close in time}
\begin{equation*}
    \mathcal{N}_{x_k}^{\delta} = \{ x_a | a \neq k \mbox{ and }  \gr{\tau}_a \in (\tau_k - \delta, \tau_k + \delta) \} ,
\end{equation*} 
where $\delta$ is a predefined threshold corresponding to time.

\begin{figure}[htp]
    \centering
    \includegraphics[width = \textwidth]{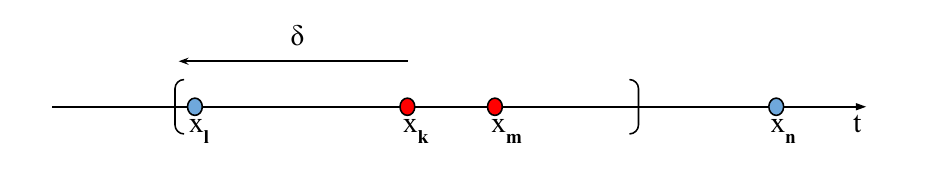}
    \vspace{-4mm}  
    \caption{Illustration of trade co-occurrence. This figure visualizes the idea of co-occurrence of trades. With a pre-defined neighbourhood size $\delta$, trade $x_l$ arrives within the $\delta$-neighbourhood of trade $x_k$, and thus they co-occur. In contrast, trade $x_n$ locates outside $x_k$'s \gr{$\delta$-}neighbourhood, and thus the two trades do not co-occur. 
    \gr{Trade $ x_n$ co-occurs with $x_m$, and both} 
trade $x_l$ and \gr{trade} $x_m$ co-occur with trade $x_k$, but they do not co-occur with each other. 
\gdr{Adapted from Figure 1 in  \cite{lu2022trade}.}
}  
    \label{fig:trade_cooccurrence}
\end{figure}

Following the definition of trade co-occurrence from \cite{lu2022trade}, we say that trade $x_k$ {\it co-occurs} with $x_l$ at level $\delta$, denoted by $x_k \overset \delta\sim x_l$, if and only if $x_l \in \mathcal{N}_{x_k}^{\delta}$. Figure \ref{fig:trade_cooccurrence} visualizes the definition, where $x_k$ co-occurs with $x_l$ and $x_m$, but does not co-occur with $x_{n}$. \gr{Trade $ x_n$ co-occurs with $x_m$, and both} 
trade $x_l$ and \gr{trade} $x_m$ co-occur with trade $x_k$, but they do not co-occur with each other. 

The co-occurrence relation is symmetric, that is, given $\delta$, if $x_k \overset \delta\sim x_l$, then $x_l \overset \delta\sim x_k$. Notice that co-occurrence of trades is not a equivalence relation. When it is clear from the context, we omit the level $\delta$ of the co-occurrence when referring to co-occurrence.  

\subsection{Pairwise co-trading score} \label{sec:co-trading_score}
Motivated by the intuition that stocks are more 
\mc{inter-dependent} 
if they are co-traded together more often, we propose pairwise co-trading scores to measure the similarity between stocks. For a pair of stocks, we calculate the similarity score by counting selected types of co-occurred trades and normalize by the total number of trades of both stocks. 

The formal definitions are as follows. \gdr{We assume a finite universe of $N$ stocks.} For stock $i$ on day $t$, the set of all trades, with direction $d^{i}\gr{\in \{ buy, sell, all \}}$, is denoted by
\begin{equation*}
    S_{t}^{i, d^{i}} = \{ x_a | \tau_a \in [t_{open}, t_{close}], s_a = i, d_a = d^{i} \},
\end{equation*}

\noindent  where 
$d^{i} = all$ denotes all trades without distinguishing between buy and sell, and $t_{open}$ and $t_{close}$ stand for the time of market open and close respectively on day $t$. Then, if given another set $S_{t}^{j, d^{j}}$, we count the number of trades for stock $j$, which co-occur with trades in $S_{t}^{i, d^{i}}$, denoted as

\begin{equation*}
    L_{t, j \rightarrow i}^{d^{j} \rightarrow d^{i}} = \sum_{x_k \in S_{t}^{i, d^{i}}} |\{x_a \in \mathcal{N}_{x_k}^{\delta} | s_{a} = j, d_{a} = d^{j}\}| ,
\end{equation*}
where $| \cdot |$ denotes \gr{the} cardinality of a set.

The pairwise co-occurrence count ind\gr{ex $c_{t,i,j}^{\delta, d^i, d^j}$} 
\gr{is a scaled count of}
the number of trades for stock $i$ with direction $d^i$, and trades for stock $j$ with direction $d^j$,  which co-occur on day $t$. \gr{Formally, it} is defined as 
\begin{equation*}
    c_{t,i,j}^{\delta, d^i, d^j} := \frac{L_{t, i \rightarrow j}^{d^{i} \rightarrow d^{j}} + L_{t, j \rightarrow i}^{d^{j} \rightarrow d^{i}} }{\sqrt{|S_{t}^{i, d^i}|} \sqrt{|S_{t}^{j, d^j}|}} .
\end{equation*} 

These pairwise co-occurrence indices have three useful properties. First, they are non-negative and higher values indicates stronger co-occurrence relations. Second, the indices are scaled, so that they can be used to compare relations across pairs of stocks. Third, the indices are symmetric. Additionally, the indices  are defined using sets of trades which are filtered based on different conditions so that they are flexible and can easily be generalized to \gr{a} customized set of orders. 

\subsection{Co-trading matrices and networks} 

Using the pairwise co-occurrence measures, we build the corresponding daily $N \times N$ co-occurrence matrix, denoted as $\mathbf{C}_{t}^{\delta, d^{i}, d^{j}}$, \gr{having entries} 
\begin{equation*}
    (\mathbf{C}_{t}^{\delta, d^{i}, d^{j}})_{i,j} = c_{t,i,j}^{\delta, d^i, d^j}.
\end{equation*}
Built from daily matrices, co-occurrence matrices over a longer time period $T$, such as months and years, are simply calculated by averaging the daily co-occurrence matrices; 
\begin{equation*}
    \mathbf{C}_{\{ T \}}^{\delta, d_{i}, d_{j}} =  \frac{1}{|T|} \sum_{t \in T} C_{t}^{\delta, d_{i}, d_{j}}.
\end{equation*}  

Taking advantage of the symmetric co-occurrence matrices, we build co-occurrence networks to represent complex structures in the stock market. We consider  dynamic co-trading networks of stocks, $G_t = (V_t, E_t)$,  $t \in T$, where each vertex $v_{i, t} \in V_t$ represents a certain stock at time $t$ and a weighted edge $e_{i,j} (t)  \in E_t$ denotes a type of co-occurrence relation between 
\gr{two} stocks at time $t$; the weight is the corresponding co-trading score. 
We use the co-trading matri\gr{ces} as representation\gr{s} of each co-trading network. The empirical study in the following sections focuses on the co-trading matrices without taking into account the directions of trades. In line with \cite{lu2022trade}, we choose $\delta = 500$ microseconds as the neighborhood size, in order to determine co-occurring trades. Therefore, we omit the superscripts for brevity and only keep the subscript for time index, e.g. $\mathbf{C}_{2017-2019}$ stands for the co-trading network aggregated from 2017 to 2019.     

\section{Empirical network analysis} \label{sec:network_analysis}

In this section, we construct a co-trading network of the whole period of study from the empirical data. We provide a visualization of the network, and show \gr{that} the co-trading matrix reflects empirical phenomen\gr{a which have been observed} 
in equity markets. Furthermore, we demonstrate that \gdr{a} co-trading 
network can capture inter-dependency of stocks beyond GICS sectors.

\subsection{Data}
The empirical studies in this paper are based on 457 stocks in US equity markets from 2017-01-03 to 2019-12-09. The \gdr{457 stocks are those } 
constituents of the Standard $\&$ Poor's ($S\&P$) 500 index 
\gdr{which have} order book and price data available over the entire period of study.

We acquire limit order book \gr{data} from the LOBSTER database (\cite{huang2011lobster}), which keeps track of submissions, cancellations and executions of limit orders for stocks traded in NASDAQ.  Each record has a timestamp with precision up to $10^{-9}$ seconds and indicates the price, size and direction of \gr{the respective} order book event. A trade occurs when a market/marketable order arrives and consumes existing limit orders, and thus can be inferred by order executions. \gr{Here we denote a trade as `buyer--initiated' if the limit order denoted a willingness to sell, and as `seller--initiated' if the limit order denoted a willingness to buy.}  To derive trades, we filter out events other than executions. Then we aggregate  records with exactly the same timestamp and direction, as they are likely to be caused by one large marketable order. The direction of a trade is opposite to those of the executed limit orders it is matched against. 
 
In addition, we obtain \mc{minutely} 
price data from LOBSTER and daily stock prices from the Center for Research in Security Prices (CRSP) database. Apart from prices, we label stock sectors using the Global Industry Classification Standard (GICS) drawn from \gdr{the} Compustat database. The GICS decomposition classifies stocks into 11 sectors of varying sizes.

\subsection{A co-trading network} \label{sec:sampel_network}
To provide a general view of co-trading, we construct a network-based matrix $\mathbf{C}_{2017-2019}$, \gr{capturing all} 
co-trading relations of the entire period of study from 2017-01-03 to 2019-12-09, without differentiating directions of trades.  

\begin{figure}[t]
    \centering
    \includegraphics[width = 1\textwidth]{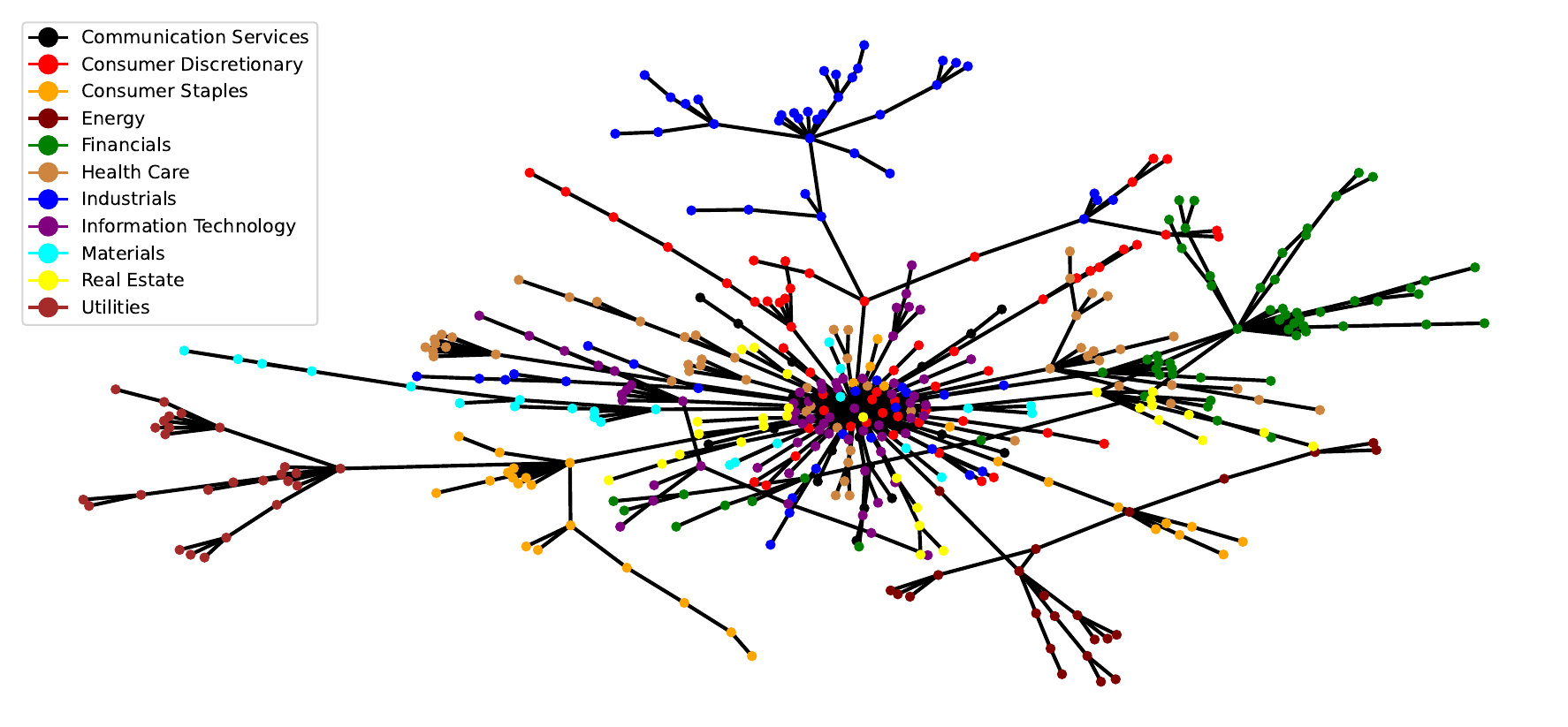}
    \caption{\small A co-trading network of the period from 2017-01-03 to 2019-12-09. \\ The figure presents the co-trading network of 457 stocks represented by the co-trading matrix, without trading directions, of the entire period of study, that is $\mathbf{C}_{2017-2019}$. Each node represents a stock, whose color indicates its GICS sector. For visualization, we use \gr{a} maximum spanning tree (\cite{mantegna1999hierarchical}) to filter out edges by maximizing sum of co-trading scores while keeping a connected network with 456 edges.}  
    \label{fig:mst_cooc_network_plot}
\end{figure}

To visualize the network, we follow \cite{mantegna1999hierarchical} 
\gdr{and} filter edges 
\gdr{using} a minimum spanning tree (MST), because the co-trading matrix is dense. To be specific, \gr{a} MST of a graph with $N$ nodes is a connected subgraph with only $N - 1$ edges \gr{such that} 
the sum of the negations of edge weights \gr{is} minimized, which can be found by Kruskal's algorithm (\cite{kruskal1956shortest}). For the implementation, we use \gdr{the Python package} `NetworkX' (\cite{hagberg2008exploring}) to filter edges and visualize networks. 
Figure \ref{fig:mst_cooc_network_plot} shows the MST of the co-trading network based on all trades of all stocks. The graph vertices represent individual stocks and their colors indicate the GICS sectors. It is noteworthy that companies within the same sector groups are often on the same branches, suggesting that stocks may frequently be co-traded in sector baskets. This \gdr{chimes}
 with the known fact that stocks in the same sector tend to move together, \mc{due to common membership in the same index traded funds (\cite{harford2005correlated}) and similar exposure to the same factors.} 
Hence, 
\gr{this} co-trading network captures meaningful patterns of the cross-sectional structure of stocks. 

Furthermore, the network visualization reveals a dense cluster at the center of the plot. The stocks close to the centroid co-trade with more stocks and are, on average, more inter-related 
with the rest of the network,  
 \gdr{compared to the stocks at the leaves of the network}.
To further investigate the cluster, we use eigenvector centrality (\cite{bonacich1972factoring}, \cite{bonacich1987power}) to measure the influence of each node on the network. For each stock $i$, 
the eigenvector centrality \gdr{is} the $i$th element of the eigenvector corresponding to the dominant eigenvalue of the co-trading matrix. \gr{The larger the centrality, the more ``influential'' the stock.}

\begin{figure}[h]
    \centering
    \includegraphics[width = 1\textwidth]{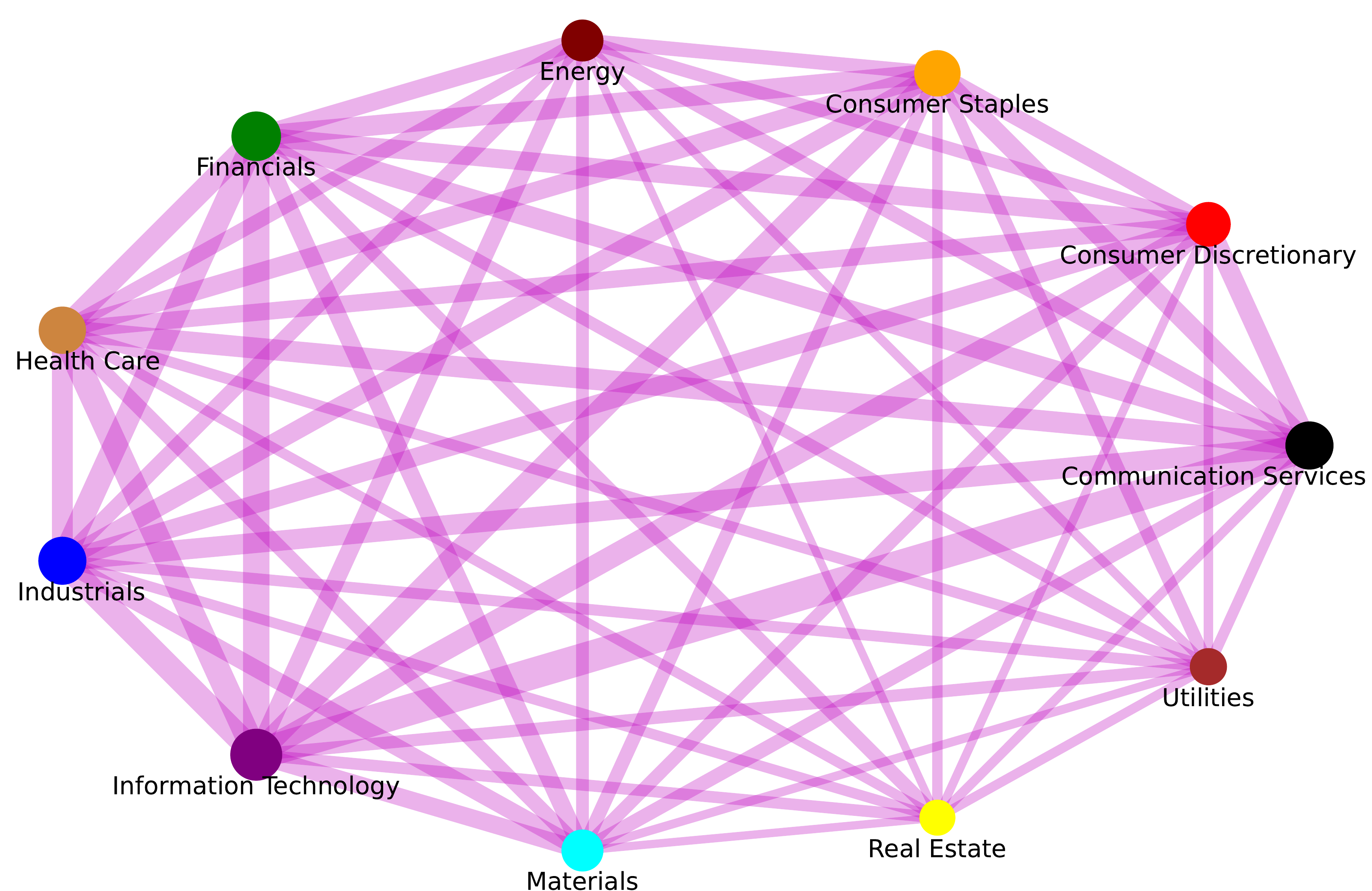}
    \caption{\small Meta-flow network of GICS sectors. \\ This figure illustrates the fully connected network of GICS sectors based 
    \gr{on the} co-trading matrix $C_{2017-2019}$. The edges are calculated by grouping stocks by their GICS sectors and averaging the co-trading scores of sector groups. }
    \label{fig:sector_meta_graph}
\end{figure}

We list \gr{the} 10 stocks with the highest \gr{eigenvector} centrality in Table \ref{tab:top_centrality_comp}. Microsoft is the most influential stock,  followed by Apple and JPMorgan Chase. Moreover, among the top 10 stocks, $60\%$ are technology companies, including Facebook, and $40\%$ are financial institutions.
 \gdr{Moreover, t}he co-trading matrix 
 \gdr{does not only model 
individual stocks, but it can}  also model relations among sectors. Following \cite{bennett2022lead}, we build a meta-flow network of sectors \gr{as follows}. We group stocks by their sector labels, \gr{use the sectors as nodes in the meta-flow network, and} use the average co-trading scores of stocks in each pair of sectors to be the sector co-trading scores\gr{, serving as edge weights in the meta-flow network}. Figure \ref{fig:sector_meta_graph} pictures the fully connected sector network, where the edge width indicates the strength of co-trading. The strongest co-trading relation is between Information Technology and Communication Services. By comparing edge widths, we can identify which sectors are more closely co-traded with a given sector. For example, \gr{the} Real Estate sector has a strong co-trading relation with \gr{the}  Financial\gr{s} sector, while \gr{it does not show a strong relationship with the other sectors.}
Moreover, we document centrality of each sector in Table \ref{tab:sector_meta_centrality}, and find  \gr{that} Information Technology, Financials and Communication Services are the most influential sectors, while Real Estate and Utilities \gr{have the lowest eigenvalue centrality}.

\begin{table}[htp]
    \caption{Top 10 stocks by eigenvector centrality. \\ This table lists 10 stocks ranked by their eigenvector centrality, as well as their company names and GICS sectors.}    
    \begin{tabularx}{\textwidth}{llll}
\toprule
Ticker   & Centrality        & Company              & Sector                 \\
\midrule
MSFT     & 0.12              & Microsoft Corp.       & Information Technology \\
AAPL     & 0.11              & Apple Inc.            & Information Technology \\
JPM      & 0.10              & JPMorgan Chase \& Co. & Financials             \\
BRK.B    & 0.10              & Berkshire Hathaway    & Financials             \\
TXN      & 0.10              & Texas Instruments     & Information Technology \\
FB       & 0.09              & Facebook Inc.        & Communication Services \\
V        & 0.09              & Visa Inc.             & Information Technology \\
AXP      & 0.09              & American Express Co   & Financials             \\
PYPL     & 0.09              & PayPal                & Information Technology \\
C     & 0.08              & Citigroup Inc.         & Financials \\
\bottomrule
\end{tabularx}

    \label{tab:top_centrality_comp}
\end{table}

\begin{table}[h]
    \caption{Centrality of sectors in \gr{the} meta-flow network. \\ This table shows the eigenvector centrality of each GICS sector in the meta-flow network. The `Rank' column reports the rank, in descending order, of each sector in terms of their centrality.}    
        \begin{tabularx}{\textwidth}{l*{3}{Y}}
        \toprule
        ~                              & Centrality        & Rank         \\ 
        \midrule
        \bftab Information Technology  & \bftab 0.38       & \bftab 1     \\ 
        \bftab Financials              & \bftab 0.35       & \bftab 2     \\ 
        \bftab Communication Services  & \bftab 0.34       & \bftab 3     \\ 
        Industrials                    & 0.33              & 4            \\ 
        Health Care                    & 0.33              & 5            \\ 
        Consumer Staples               & 0.32              & 6            \\ 
        Consumer Discretionary         & 0.29              & 7            \\ 
        Energy                         & 0.26              & 8            \\ 
        Materials                      & 0.26              & 9            \\ 
        Utilities                      & 0.21              & 10           \\         
        Real Estate                    & 0.20              & 11           \\ 
        \bottomrule
    \end{tabularx}
    \label{tab:sector_meta_centrality}
\end{table}

\subsection{Clustering analysis} \label{sec:static_clustering_analysis}
So far, we have observed the presence of clusters, or communities, in the co-trading network of stocks. In the following part, we introduce an unsupervised clustering method to classify every stock,
with the aim to group similar stocks into the same cluster. By conducting clustering analysis on the network of $\mathbf{C}_{2017-2019}$ in Figure \ref{fig:mst_cooc_network_plot}, we show that the co-trading matrix not only captures sector structure, but also incorporates associations beyond sectors. 

\subsubsection{Methodology and evaluation metrics}  \label{cluster_method}
Spectral clustering is a family of algorithms (\cite{shi2000normalized, ng2002spectral, cucuringu2019sponge, cucuringu2020hermitian}), built upon spectral graph theory, to detect communities or clusters in networks. For details, \cite{von2007tutorial} provides a comprehensive survey on spectral methods and their theoretical backgrounds. Given a co-trading matrix, we apply a spectral clustering algorithm, outlined in Appendix \ref{appendix_spectral_clustering}, to identify clusters in our universe of stocks. The number of clusters is a hyper-parameter of this algorithm, which we need to determine beforehand. Although the spectral clustering depends on \gdr{a} random initialization, in Appendix \ref{appendix_clustering_initialization}, we show that the algorithm is robust in our empirical setting.

\gr{The} Adjusted Rand index (ARI) (\cite{hubert1985comparing}) is our measure of similarity between clusters. \mc{The valid range of ARI is $[\gdr{-1}, 1]$, where 1 is achieved if and only if two clusters perfectly match. Notice that ARI can take negative values, which means the two clusters are not similar at all. In general, a higher ARI indicates that two clusters are more similar to each other, and vice versa. A value close to 0 indicates that points are assigned into clusters \gdr{almost} randomly.}

\subsubsection{Data-drive clusters v.s. GICS sectors}
We apply the spectral clustering method on the co-trading network By analyzing the clusters, we discover that co-trading networks capture sectors in the US stock market, with an ARI of 0.56, which supports our observations in \Cref{sec:sampel_network}.

\begin{figure}[h] 
    \centering
    \includegraphics[width = 1\textwidth]{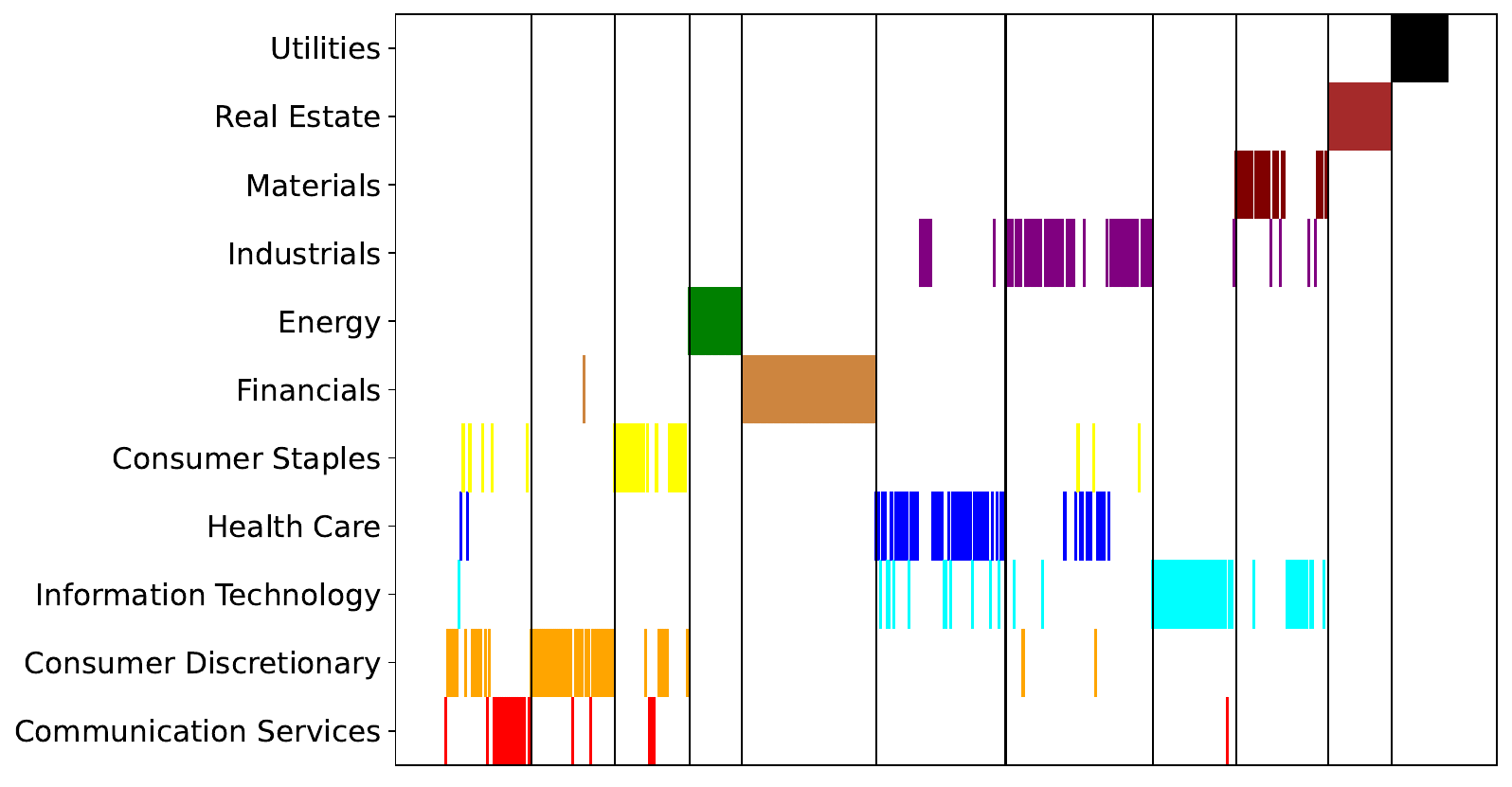}
    \caption{\small Clusters v.s. sectors. \\ This plot visualizes the overlap between the data-driven clusters, derived from $\mathbf{C}_{2017-2019}$ by spectral clustering, and the GICS sectors. The horizontal axis indexes stocks grouped by data-driven clusters, separated by vertical lines. The vertical axis indicates GICS sector labels. \mc{The colored area of a sector is indicative of its size.} } 
    \label{fig:cooc_clusters_plot}
\end{figure}

Figure \ref{fig:cooc_clusters_plot} shows the commonality between GICS sectors and clusters detected in the co-occurrence network corresponding to \gr{the} matrix $\mathbf{C}_{2017-2019}$. By setting the number of clusters to 11, which is also the number of GICS sectors after \gr{the} classification change in 2018, we 
observe that the unsupervised clustering method can recover the sector groups to a good standard. This is especially the case for companies in the  Financial, Utilities, Real Estate and Energy sectors, when there are hardly any mismatches. There are also clear clusters for stocks in \gr{the sectors} Materials, Health Care, Industrials and Consumer Staples, 
with small amounts of disagreements. Comparatively, the structure of the  Information Technology sector is strikingly different, with one major cluster containing most of the stocks and most of the other stocks well spread out across other clusters. Moreover, the majority of stocks from the Consumer Discretionary \gr{sector} concentrate in one cluster, and others are grouped with stocks in the Consumer Staples and Communication Services sectors.

\section{Dynamics of co-trading networks and clusters} \label{sec:temporal_networks}
In addition to describing the long-term structure of the US equity market, time series of co-trading matrices built at higher frequencies contain information on the dynamic of topological structures of the market. We observe that co-trading across GICS sectors grows over time. To further explore, we compare the daily ARIs between generated clusters and GICS sectors and find a downward trend in their similarity. Moreover, we investigate the temporal changes of daily clusters and detect regime switching over the period of study.

\subsection{Temporal evolution of co-trading networks} \label{sec:daily_networks}
\begin{figure}[htp]
    \begin{subfigure}{\textwidth}
    \begin{subfigure}{0.32\textwidth}
    \includegraphics[width = \textwidth, height = \textwidth]{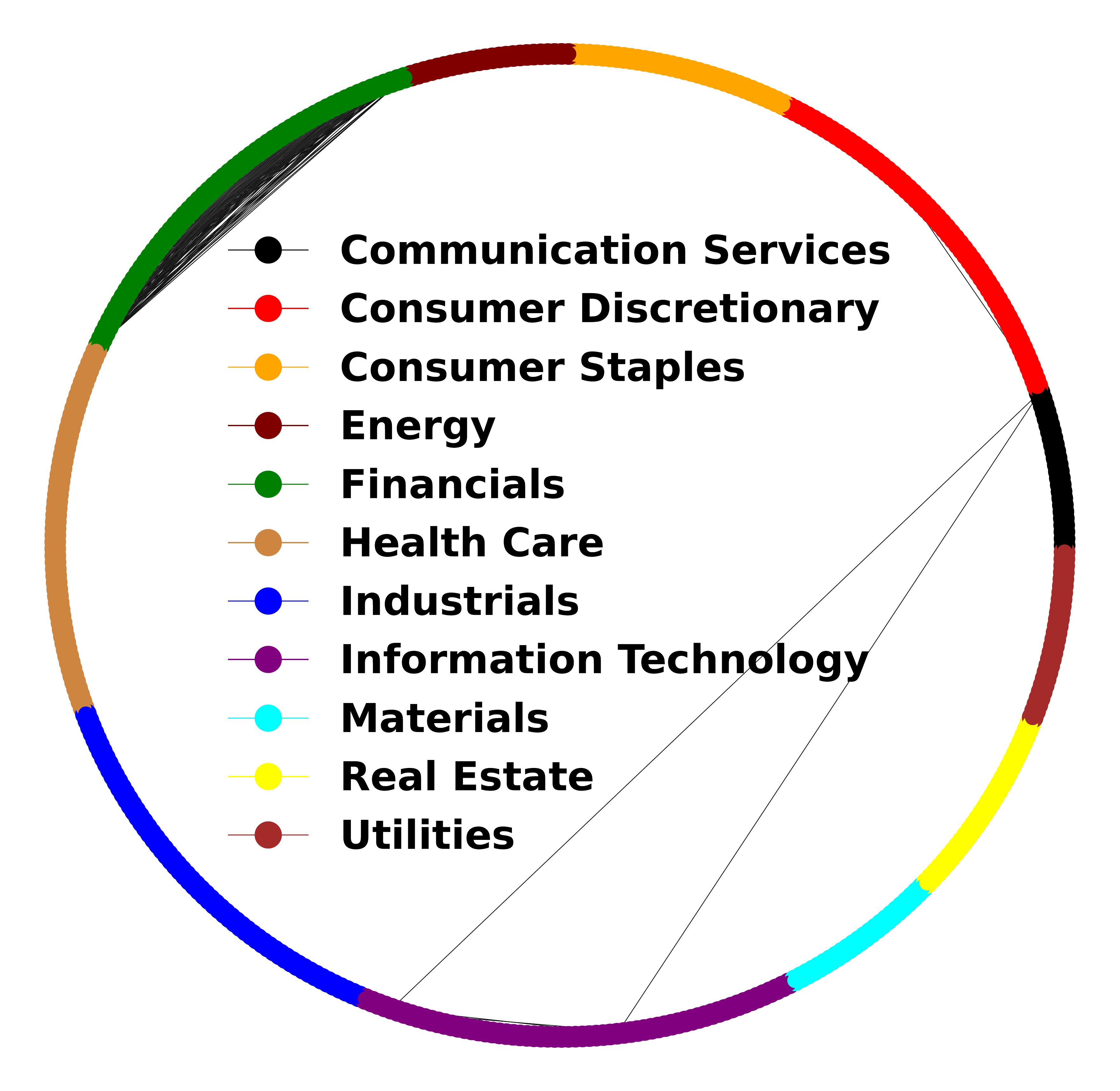}
    \end{subfigure}
    \hspace{\fill}
    \begin{subfigure}{0.32\textwidth}
    \includegraphics[width = \textwidth, height = \textwidth]{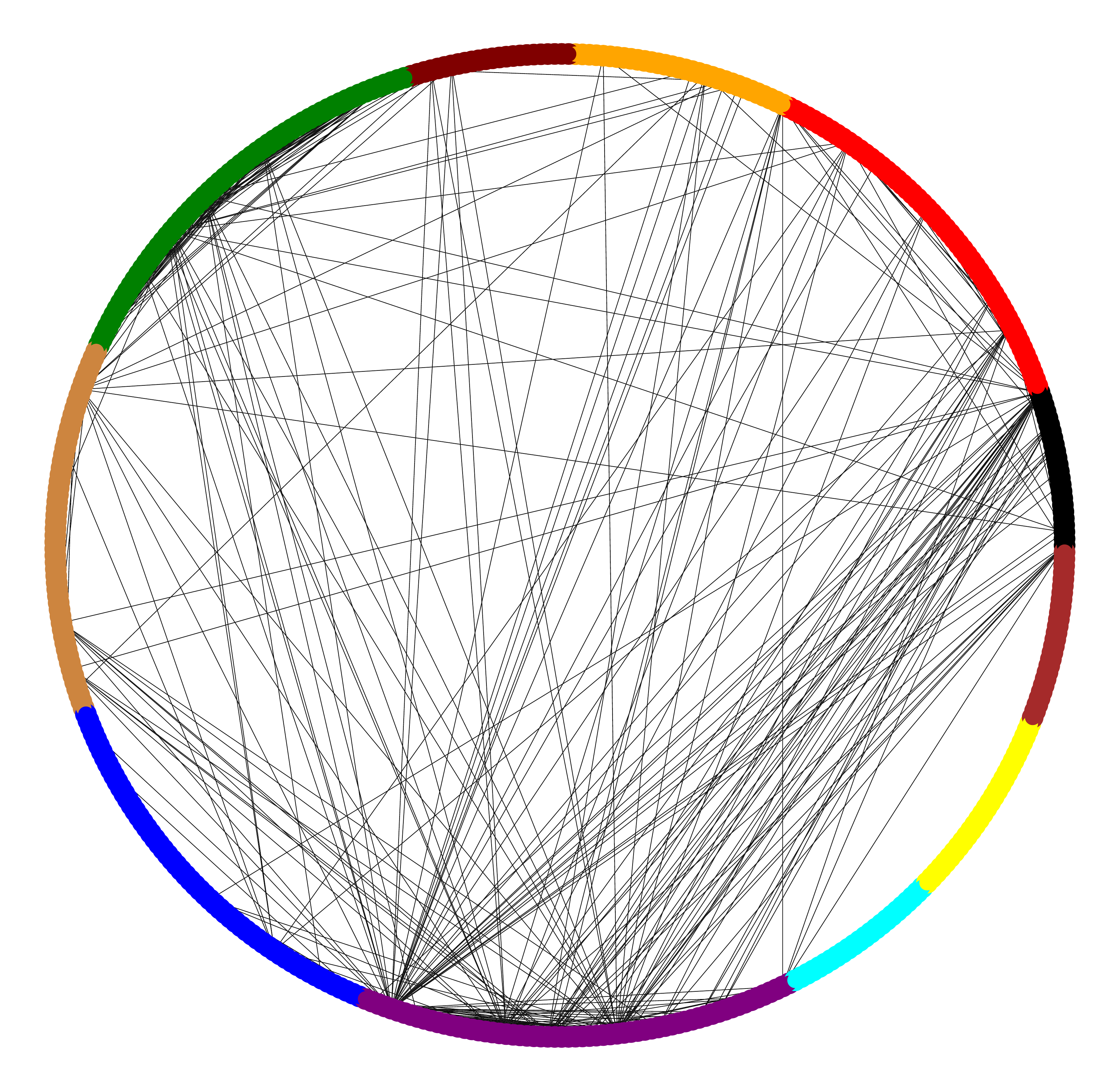}
    \end{subfigure}
    \hspace{\fill}
    \begin{subfigure}{0.32\textwidth}
    \includegraphics[width = \textwidth, height = \textwidth]
    {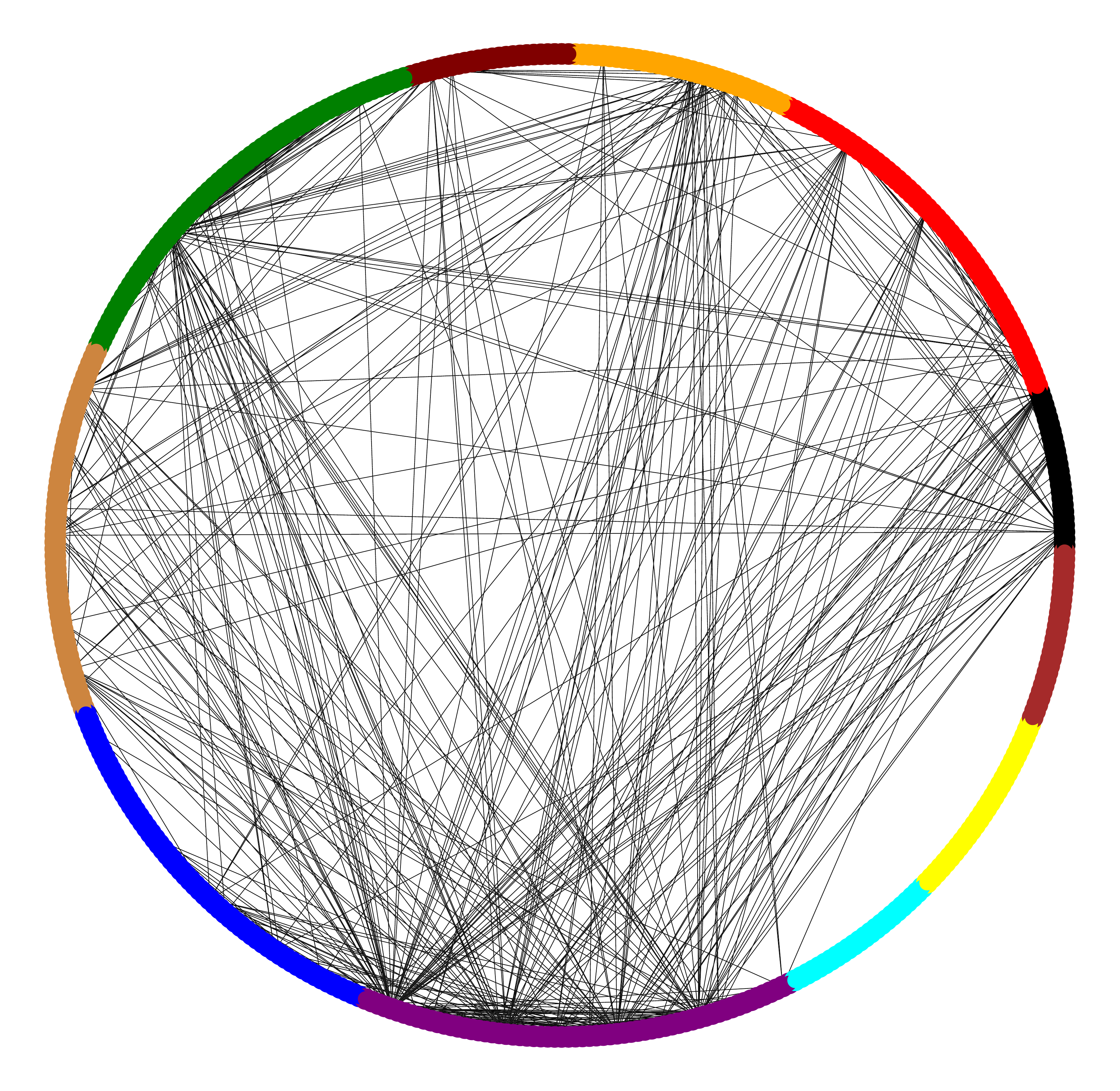}
    \end{subfigure}
    \end{subfigure}

    \begin{subfigure}{\textwidth}
    \begin{subfigure}{0.32\textwidth}
    \includegraphics[width = \textwidth, height = \textwidth]{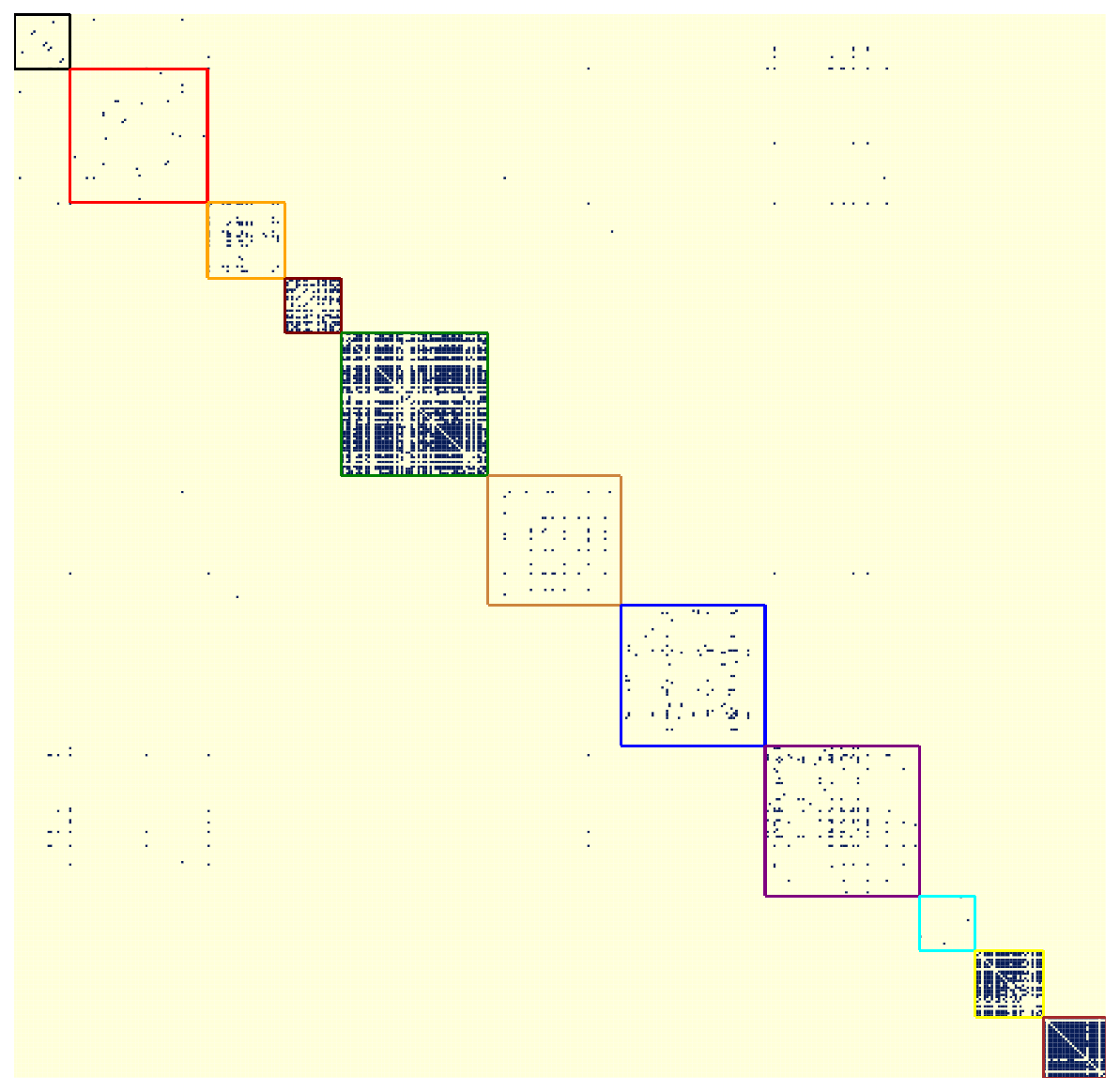}
    \caption{\small 2017-01}
    \end{subfigure}
    \hspace{\fill}
    \begin{subfigure}{0.32\textwidth}
    \includegraphics[width = \textwidth, height = \textwidth]{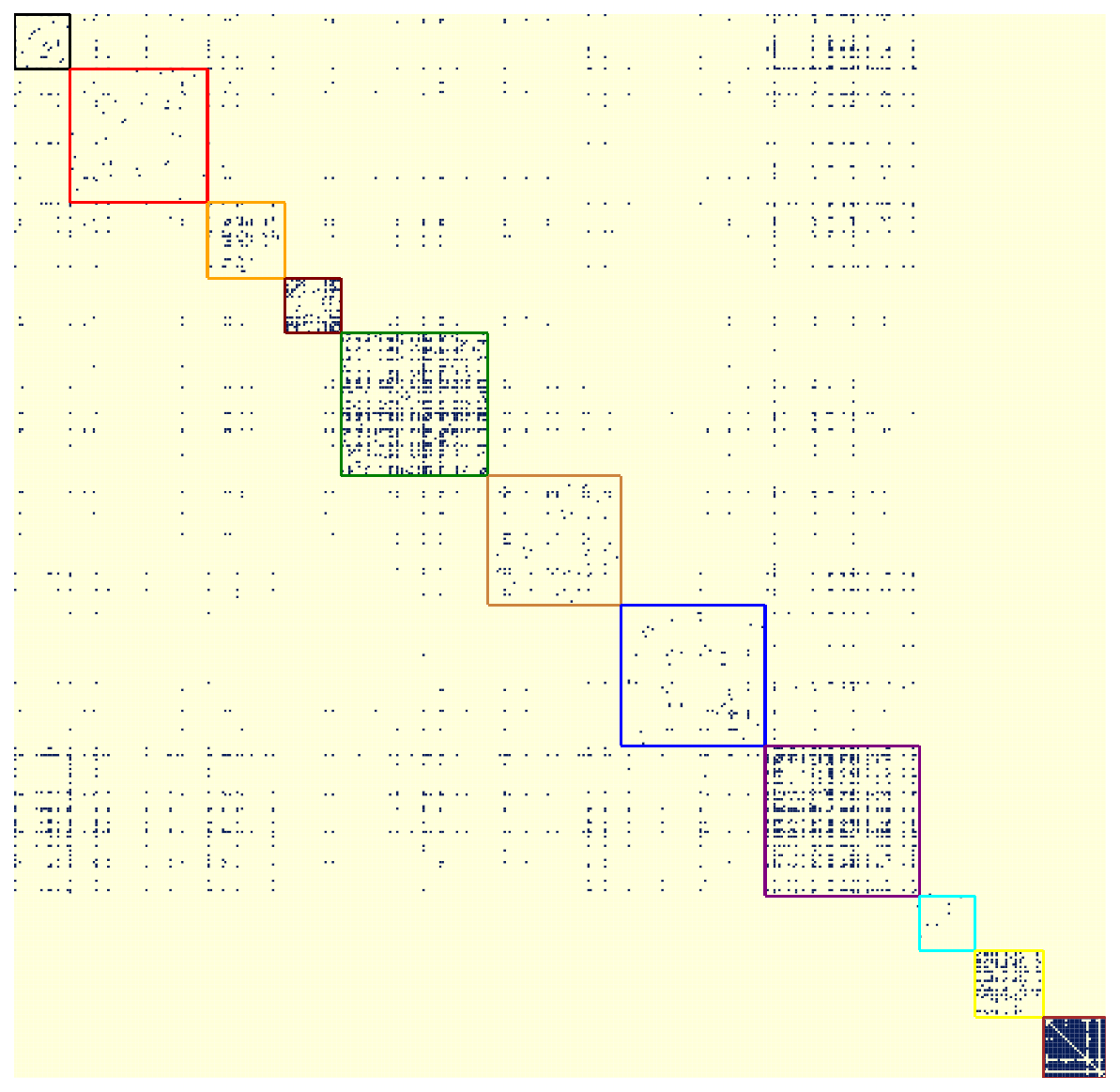}
    \caption{\small 2018-01}
    \end{subfigure}
    \hspace{\fill}
    \begin{subfigure}{0.32\textwidth}
    \includegraphics[width = \textwidth, height = \textwidth]
    {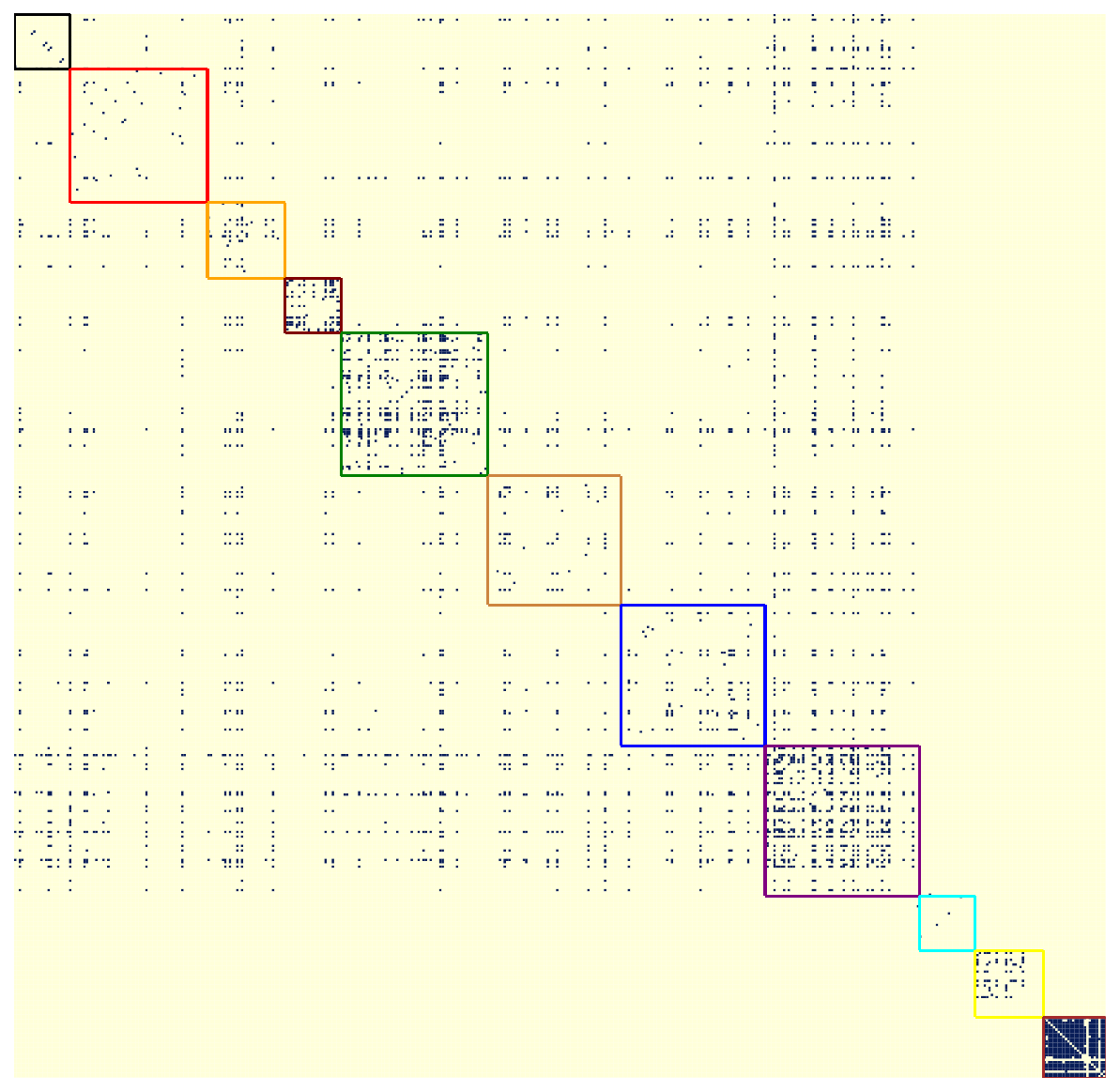}
    \caption{\small 2019-01}
    \end{subfigure}
    \end{subfigure}

    \caption{\small Thresholded temporal co-trading networks. \\ These three networks are based on co-occurrence of all trades regardless of trading directions. For visualization, we threshold on co-trading scores and only keep top $1 \%$ of edges. The first row contains the thresholded networks and the second row includes heatmaps of the corresponding co-trading matrices. From left to right, are networks and co-trading matrices of 2017-01, 2018-01 and 2019-01. Colors stand for the GICS sectors. } \label{fig:threshold_cooc_net} 
\end{figure}

In Figure \ref{fig:threshold_cooc_net}, we plot  co-trading networks and their corresponding adjacency matrices, for the month of January, for the years 2017, 2018 and 2019, with colors representing the GICS sectors. After thresholding (\cite{huang2009network}), we only retain $1 \%$ of the edges, \gr{selected such that we preserve the edges} with the highest weight. \gdr{We find that t}he co-trading structures in the US equity market gradually change over the sample period. At the start of the period, the within-sector interactions are prominent, with only a few strong co-trading relations across the GICS sectors. However, towards the end of the period, an increasing number of edges connect stocks from different sectors. The series of heatmaps clearly shows a  pattern of decreasing within-sector co-trading connections, such as Financials, Real Estate, \gdr{and} Energy. 
In contrast, the strong co-trading connections of constituents of the Information Technology sector grow both within and across sectors. Co-trading relations among stocks in the Utility sector remain strong and isolated from other sectors over time. Therefore, the sets of stocks that speculators and investors tend to trade together in their portfolios are actually time-varying and deviate gradually from the standard GICS sectors. These findings also provide supporting evidence for the importance of dynamic data-driven clustering, that goes beyond the usual sector-based decomposition.

\subsection{Temporal clustering analysis} \label{sec:daily_clustering}

\mc{Similar to uncovering
long-term agreement between data-driven clusters and GICS sectors in \Cref{sec:static_clustering_analysis}, we are also interested in assessing and quantifying the extent to which the cross-sectional} trading behaviors deviate from sectors within a short period of time. Therefore, we also proceed with clustering stocks on a daily basis, and then compare the similarity between clusters and sectors, via the same \gdr{ARI measure}.
Since the true number of communities in the stock market is \mc{unknown},
we consider multiple values \gr{for the number} of clusters, 
\gr{namely} 11 (\gr{the} same \gr{number} as \gr{the number of} GICS \gr{sectors}), 15, 20 and 50. 

\begin{figure}[htp]
    \centering
    \includegraphics[width = \textwidth]{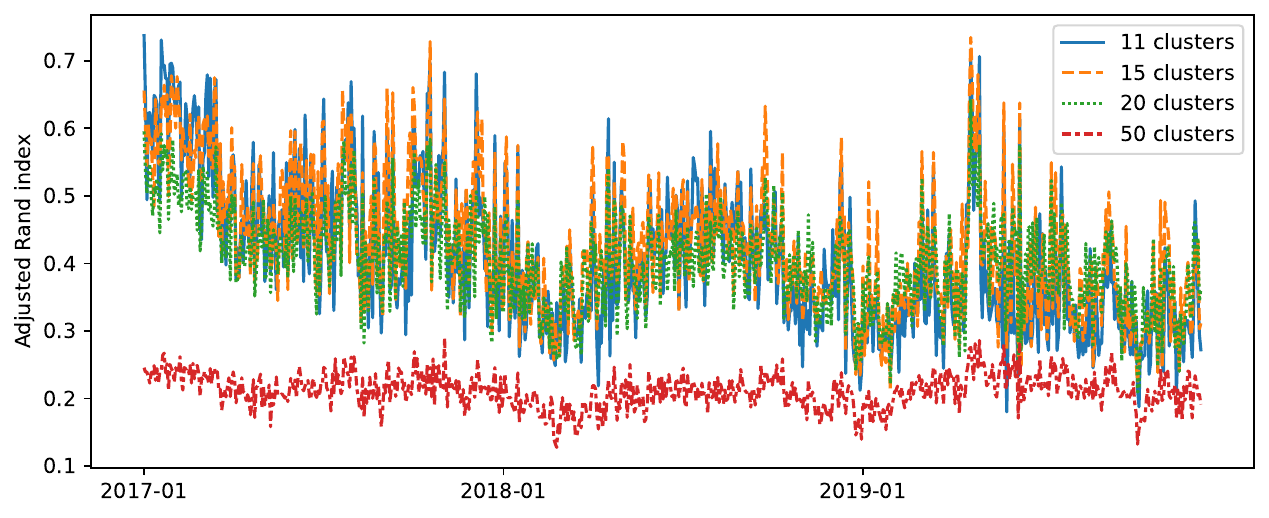}
    \caption{\small 
    \gdr{ARI} between daily clusters and GICS sectors. \\ This figure sketches dynamics similarity between data-driven clusters and GICS sectors. \gr{For e}very day, we derive data-driven clusters and calculate the adjusted Rand index with GICS, with predefined numbers of clusters  $\{11, 15, 20, 50\}$. We then plot the ARIs over time from 2017-01-03 to 2019-12-09, 
    \gr{so that} each line represents one \gr{choice} 
    of \gr{number of} clusters. 
    }
    \label{fig:sector_cluster_ari}
\end{figure}

\begin{table}[htp]
    \caption{Statistics of daily ARI. \\ This table shows summary statistics of daily ARIs between dynamic clusters and fixed GICS sectors plotted in Figure \ref{fig:sector_cluster_ari}. For different number of clusters, the table reports the mean and standard deviation of ARIs from 2017-01-03 to 2019-12-09, \gr{as well as the signal-to-noise ratio}.}    
    \begin{tabularx}{\textwidth}{l*{5}{Y}}
\toprule
                   & \multicolumn{4}{c}{Clusters} \\
\cmidrule(lr){2-5}
                   & 11    & 15    & 20    & 50   \\
\midrule
Mean               & 0.41  & 0.43  & 0.40  & 0.21 \\
Standard deviation & 0.11  & 0.10  & 0.07  & 0.03 \\
\gr{Signal-to-noise ratio} &
3.73& 4.30 & 5.71 & 7.00  \\ 
\bottomrule
\end{tabularx}

    \label{tab:stats_ari}
\end{table}  

Figure \ref{fig:sector_cluster_ari} 
\gr{shows the} time series of ARIs \gr{between daily clusters and GICS sectors}. 
For \gr{a} small number of clusters, we observe meaningful levels of similarity between the data-driven clusters and economic sectors. In contrast, when the number is large as 50, the similarities are constantly low. As empirical 
\gr{observations indicate that} 
market participants tend to trade sectors, we conclude that \gr{the} true number of clusters is relatively small. Moreover, in Table \ref{tab:stats_ari} we summarize the mean and standard deviation of daily ARIs, for each choice of cluster numbers, over the entire period, \gr{echoing} 
our visual findings. In addition, although the average ARI values of 11, 15 and 50 clusters are comparable, the variation drops by 0.03 as the number of clusters increases from 15 to 20, which is more conspicuous than 
\gr{the increase} from 11 to 15. \gr{Based on the signal-to-noise ratio and taking the ARI into account,} 
\mc{we settle to further investigate the case of 20 clusters}. 

Focusing on the temporal ARI curve of 20 clusters, we 
highlight two empirical findings. Firstly, at the daily frequency, co-trading behaviors align well with economic sectors, with frequent fluctuations. Secondly, we observe a downward trend in similarities, hinting that the sector structures become less prominent from 2017 to 2019. This reinforces the observation from Figure \ref{fig:threshold_cooc_net} concerning 
the growth of strong co-trading relations across GICS sectors. 
As expected, our co-trading networks embed dynamic structures in stock markets beyond static economic sectors.   

\subsection{Temporal stability of clusters and regime detection}

\begin{figure}[htp]
    \centering
    \includegraphics[width = 0.5\textwidth]{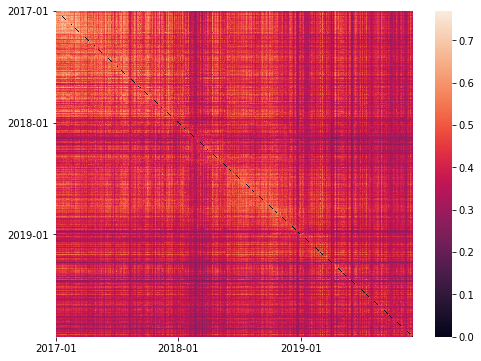}
    \caption{\small Similarity of daily clusters. \\ This heatmap shows the similarity between days. \gdr{For e}very day, we group stocks into 20 clusters by applying spectral clustering on the daily co-trading matrix. Then, for each pair of days, we calculate \gr{the} ARI between the clusters. }
    \label{fig:temporal_ari_heatmap}
\end{figure}

\begin{figure}[htp]
    \centering
    \includegraphics[width = 1\textwidth]{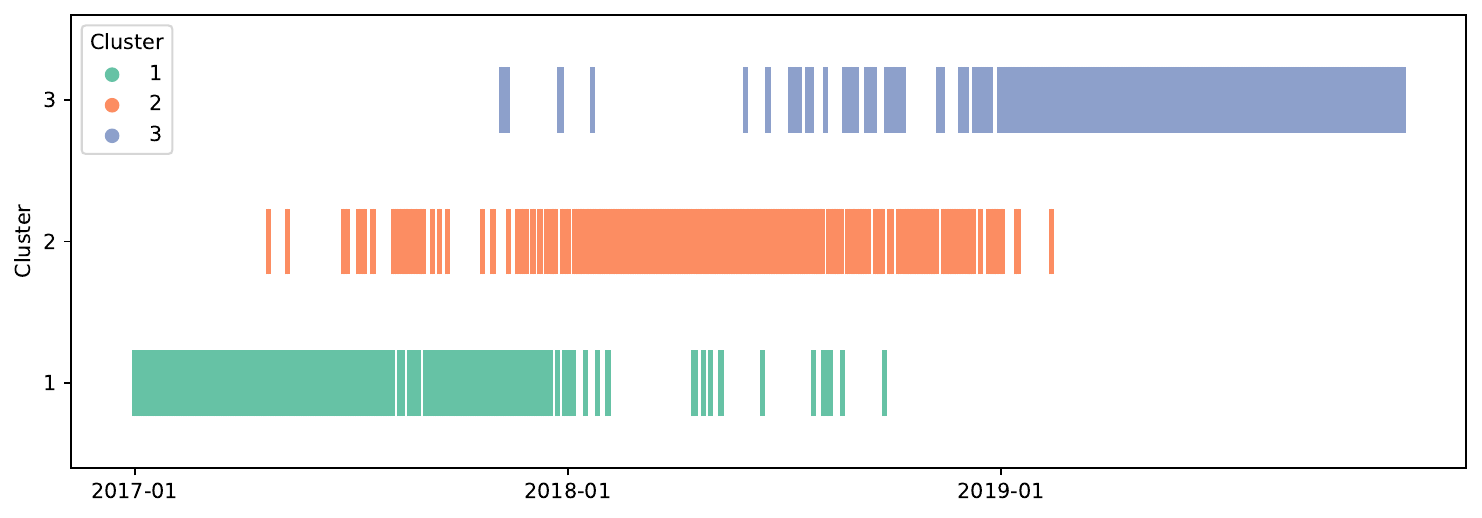}
    \caption{\small Regimes detected by spectral clustering. \\ This figure shows the clusters of days, derived by applying spectral clustering on the heatmap of ARI in Figure \ref{fig:temporal_ari_heatmap} to classify days into three regimes. The horizontal axis indicates dates and the vertical axis indicates which cluster each day belongs to.}
    \label{fig:temporal_clusters_plot}
\end{figure}

\gr{In addition to} comparing daily clusters with the benchmark, we 
also measure the similarity between clusters corresponding to every pair of days. 
 Figure \ref{fig:temporal_ari_heatmap} \gr{shows a heatmap  of all such} pairwise ARIs, based on 20 clusters. It is noteworthy that the colors along the diagonal get darker from upper left to lower right corner, which 
 \gr{indicates} that the market structures become unstable \gr{over} 
 time.
 \gr{Moreover,} 
 we distinguish three blocks \mc{along the main} diagonal. For \gr{a} quantitative regime detection, we apply the spectral clustering method \update{(\Cref{appendix_spectral_clustering})} on the heatmap, and clearly identify three clusters of trading days, \gr{shown} in Figure \ref{fig:temporal_clusters_plot}.

\section{Co-trading and covariance} \label{sec:co-trading_covariance}
In this section, we 
connect the co-trading behavior to co-movement in stock prices. A direct measure of price co-movement is the covariance matrix. On every day $t$, we 
define a {\it realized covariance matrix}, denoted as $\Hat{\mathbf{\Sigma}}^{R}_{t}$, as a proxy of the true covariance $\mathbf{\Sigma}_{t}$, based on logarithmic returns of stocks. 
We conduct a regression analysis and 
\gdr{uncover} significant and positive associations between co-trading and realized covariance matrices. \update{Furthermore, we find that co-trading matrices significantly explain the idiosyncratic covariance among stocks.}
 
\subsection{The realized covariance matrix and \update{factor-based decomposition}}
For each day $t$, we define the \textit{realized covariance matrix} following the three steps below. Since we only use intraday data, the index $t$ is omitted in this part to avoid ambiguity.
Firstly, we split the normal trading period from 9:30 to 16:00 into \gdr{$m$} equally spaced and non-overlapping intervals of length $\Delta$. We denote $Int = \{\tau_1, \tau_2, \dots, \tau_m\}$ as the set of left end-points of each interval.
Secondly, we define the logarithmic return of all stocks, $\mathbf{r}_{\tau} \in \mathbb{R}^{N}$, for each period, indexed by the left point of the interval $\tau$, as
\begin{equation*}
    \mathbf{r}_{\tau} = \log(\mathbf{P}_{\tau}) - \log(\mathbf{P}_{\tau - \Delta}),
\end{equation*}
where $\mathbf{P}_{\tau} \in \mathbb{R}^{N}$ are mid prices at time $\tau$. 
Finally, \mc{we build the 
realized covariance matrix as}
\begin{equation*}
    \Hat{\mathbf{\Sigma}}^{R} = \sum_{\tau \in Int}^{} \mathbf{r}_{\tau}^{} \mathbf{r}_{\tau}^{T}.
\end{equation*}

For the empirical study in the following sections, we construct daily realized covariance matrices with sampling frequency of 
$\Delta$ = 5 min, in line with \cite{andersen2001distribution,liu2015does}.

\update{Furthermore, assuming that the intraday logarithmic stock returns, $\mathbf{r}_{\tau} \in \mathbb{R}^{N}$, at time $\tau$, follow a linear $K$-factor model, they can be decomposed as
\begin{equation*}
    \mathbf{r}_{\tau} = \beta \mathbf{f}_{\tau} + \mathbf{u}_{\tau},
\end{equation*}
where $\mathbf{f}_{\tau} \in \mathbb{R}^{K}$ is a vector of latent factors, $\beta \in \mathbb{R}^{N \times K}$ is a static loading matrix, and $\mathbf{u}_{\tau} \in \mathbb{R}^{K}$ is the idiosyncratic component which is assumed to be independent of $\mathbf{f}_{\tau}$. Therefore, the covariance matrix of $\mathbf{r}_\tau$ can be decomposed as 
\begin{equation} \label{eq:decompose_covariance}
    \mathbf{\Sigma} = \beta \mathbf{\Sigma}^{f} \beta^{T} + \mathbf{\Sigma}^{u},
\end{equation}
where $\mathbf{\Sigma}^{f} = \sum_{\tau \in Int} \mathbf{f}_{\tau}^{} \mathbf{f}_{\tau}^{T}$ is the low-rank covariance matrix of factors evaluated on the left endpoints of the intervals, and $\mathbf{\Sigma}^{u} = \sum_{\tau \in Int} \mathbf{u}_{\tau}^{} \mathbf{u}_{\tau}^{T} $ is the idiosyncratic covariance. 
\cite{ait2017using} prove that, under 
additional assumptions, one can use the eigenvalues and eigenvectors to approximate the factor and idiosyncratic covariance matrices,  and that the approximation errors are bounded. That is, one may write 
\begin{equation*}
    \begin{aligned}
        \beta \mathbf{\Sigma}^{f} \beta^{T} & \approx \sum_{k=1}^{K} \lambda_{k}^{} \mathbf{v}_{k}^{} \mathbf{v}_{k}^{T}, \\
        \mathbf{\Sigma}^{u} & \approx \sum_{k=K+1}^{N} \lambda_{k}^{} \mathbf{v}_{k}^{} \mathbf{v}_{k}^{T}, 
    \end{aligned}
\end{equation*}
where $\lambda_{k}$ is the $k$-th largest eigenvalue of $\mathbf{\Sigma}$ and $\mathbf{v}_{i}$ denotes its corresponding eigenvector. 
Thus, we can write the sample covariance matrix as 
\begin{equation}
    \Hat{\mathbf{\Sigma}}^{R} = \sum_{k=1}^{K} \Hat{\lambda}_{k}^{} \Hat{\mathbf{v}}_{k}^{} \Hat{\mathbf{v}}_{k}^{T} + \Hat{\mathbf{\Sigma}}^{u},
\end{equation}
where $\Hat{\lambda}_{k}^{}$ and $\Hat{\mathbf{v}}_{k}^{}$ are the $k$-th largest eigenvalue and eigenvector of the sample covariance matrix, and $\Hat{\mathbf{\Sigma}}^{u} = \sum_{k=K+1}^{N} \Hat{\lambda}_{k}^{} \Hat{\mathbf{v}}_{k}^{} \Hat{\mathbf{v}}_{k}^{T}$ is the approximation of the idiosyncratic covariance. In the following empirical studies, we 
adopt four values of the number of latent factors, $K \in \{1, 3, 5, 10 \}$.}

\subsection{Network regression analysis}

In order to \gr{assess significance of the relationships} 
between co-trading and covariance matrices, we perform a network regression on each day $t$
\begin{equation} \label{reg_qap}
     \Hat{\mathbf{\Sigma}}^{R}_{t} = \alpha_{t} \mathbf{1} + \gamma_{t} \mathbf{C}_{t} + \mathbf{E}_{t},
\end{equation}
where $\alpha_{t} \in \mathbb{R}$ is the intercept term, $\mathbf{1} \in \mathbb{R}^{N \times N}$ is an all-ones matrix,  $\gamma_{t} \in \mathbb{R}$ is the regression coefficient, and $\mathbf{E}_{t} \in \mathbb{R}^{N \times N}$ is the residual matrix. 
\gr{As} there are cross-sectional autocorrelations among stocks, 
it is \gr{not} appropriate to assume independence in the residuals. \gr{Instead, t}o draw inference on the regression coefficients, we use \gr{the} quadratic assignment procedure (QAP) (\cite{mantel1967detection}, \cite{krackardt1987qap}, \cite{krackhardt1988predicting}). \gdr{This is a permutation test for comparing an observed matrix and an explanatory matrix; in the explanatory matrix, the order of the stocks is permuted, so that any correlation between the stocks in  the two matrices is destroyed.} 
We implement this and the 
\gdr{other} tests in this section using the R package `asnipe' (\cite{farine2013animal}). Table \ref{tab:qap} reports the regression coefficients. Positive relations between covariance and co-trading matrices appear in all days over the sample period, with mean and median of daily regression coefficients equal 5.10 and 4.48, respectively. Moreover, $98.51 \%$ of these positive relations are statistically significant at the $5 \%$ significance level. 

\begin{table}[htp]
    \caption{Simple network regression. \\ This table reports the mean, median and standard deviation of daily network regression coefficients in \eqref{reg_qap}, as well as their $p$-values. \gr{The $p$-values are obtained using the QAP with 2000 simulations.} We run one regression for each trading day from 2017-01-03 to 2019-12-09. 
    }    
    \begin{tabularx}{\textwidth}{l*{3}{Y}}
\toprule
     & $\gamma_{t}$ & $p$-value  \\
\midrule
Mean &    5.10          &     0.005     \\
Median &    4.48        &     0.001     \\
Standard deviation &    2.86          &     0.034     \\
Percentage positive &   100     & -   \\    
Percentage significant &    98.51 & -  \\
\bottomrule
\end{tabularx}
    \label{tab:qap}
\end{table}

In the next step, we \gr{assess whether} 
co-trading networks explain cross-sectional variation in covariance beyond GICS sectors. To account for the sector structure in \gr{the} regression, we introduce a static network of GICS sectors, as follows
\begin{equation*}
    \mathbf{S}_{i,j} = 
    \begin{cases}
        1, & \text{if stock $i$ and stock $j$ belong to the same cluster}  \\
        0, & \text{otherwise}.
    \end{cases}
\end{equation*}
Then, controlling for the sector network, we perform the following regression 
\begin{equation} \label{reg_mrqap}
     \Hat{\mathbf{\Sigma}}^{R}_{t} = \alpha_{t} \mathbf{1} + \gamma_{t}^{C} \mathbf{C}_{t} + \gamma_{t}^{S} \mathbf{S} + \mathbf{E}_{t},
\end{equation}
where $\alpha_{t}, \gamma_{t}^{C}, \gamma_{t}^{S} \in \mathbb{R}$ are the intercept and regression coefficients. We conduct \gr{a} QAP in a multiple regression setting (MRQAP) (\cite{krackhardt1988predicting}) using double semi-partialing (SDP) (\cite{dekker2007sensitivity}). 

\begin{table}[htp]
    \caption{Multiple network regression, realized covariance v.s. co-trading. \\ This table reports the mean, median and standard deviation of daily network regression coefficients in \eqref{reg_mrqap}, as well as their $p$-values. We perform one regression for each trading day from 2017-01-03 to 2019-12-09. For inference, we use MRQAP with SDP, \mc{in line with \cite{dekker2007sensitivity}}  and simulate 2000 runs to calculate $p$-values.}    
    \begin{tabularx}{\textwidth}{l*{3}{Y}}
\toprule
     & $\gamma_{t}^{C}$ & $p$-value   \\
\midrule
Mean &    3.61          &     0.030     \\
Median &    3.04        &     0.001     \\
Standard deviation &    2.52          &    0.116     \\
Percentage positive &   99.73     & -   \\    
Percentage significant &    89.81 & -   \\
\bottomrule
\end{tabularx}

    \label{tab:mrqap}
\end{table}

The regression coefficients for the both 
co-trading and sector networks are \gr{shown} in Table \ref{tab:mrqap}; \gr{they are highly significant, indicating that there is a relationship between the co-trading networks and the realized  covariance matrix}. It is noteworthy that the conclusion holds even \gr{when accounting for} 
sectors. The mean and median of daily $\gamma_{t}^{C}$ are 3.61 and 3.04, with $99.73 \%$ of positive values. During the whole sample period, $89.81 \%$ of the coefficients are statistically significant. 
As expected, sector networks are also positively correlated with price co-movements and significantly associated to the covariance structure of the stocks. With GICS sectors included in regressions, the coefficients of daily co-trading matrices are lower, since co-trading matrices evidently incorporate sector structures as well. 
\gr{Still, even then} co-trading matrices \update{have explanatory power on} covariance of stocks beyond GICS sectors.

\update{
Furthermore, we \gdr{find} 
that the co-trading matrix captures 
\gdr{some} patterns in idiosyncratic covariance among stocks, beyond the common factors. We perform 
\gdr{a} factor-based covariance decomposition and regress idiosyncratic components against co-trading matrices with control for the sector network, as follows

\begin{equation} \label{reg_residual_mrqap}
     \Hat{\mathbf{\Sigma}}^{u}_{t} = \alpha_{t} \mathbf{1} + \gamma_{t}^{C} \mathbf{C}_{t} + \gamma_{t}^{S} \mathbf{S} + \mathbf{E}_{t},
\end{equation}
where $\alpha_{t}, \gamma_{t}^{C}, \gamma_{t}^{S} \in \mathbb{R}$ are the intercept and regression coefficients.

We report the results in Table \ref{tab:mrqap_residual}. For the  $K$-factor model with $K=1$, we remove the covariance on the direction of the first principal component from the realized covariance. Compared with Table \ref{tab:mrqap}, the average $p$-value decreases from 0.030 to 0.020, and the percentage of significant coefficients increases from 89.81\% to 95.24\%. Moreover, by separating \mcc{the} covariance of more latent factors, the coefficients become positive and significant over the whole period. Therefore, the co-trading behaviors have explanatory power on the covariance of equity returns, beyond the price co-movements driven by common risk factors.  

\begin{table}[htp]
    \caption{Multiple network regression, idiosyncratic covariance v.s. co-trading. \\ This table reports the mean, median and standard deviation of daily network regression coefficients in \eqref{reg_residual_mrqap}, as well as their $p$-values. The `Factor' column indicates the number of latent factors for covariance decomposition. We perform one regression for each trading day from 2017-01-03 to 2019-12-09. For inference, we use MRQAP with SDP, \mc{in line with \cite{dekker2007sensitivity}}  and simulate 2000 runs to calculate $p$-values.}

\begin{tabularx}{\textwidth}{l*{4}{Y}}
\toprule
Factor   & & $\gamma_{t}^{C}$ & $p$-value     \\
\midrule
\multirow{5}{*}{1} &       Mean &          2.57 &     0.020 \\
       &                 Median &          2.16 &     0.001 \\
       &     Standard deviation &          1.73 &     0.102 \\
       &    Percentage positive &         99.46 &         - \\
       & Percentage significant &         95.24 &         - \\

\midrule
\multirow{5}{*}{3} &                   Mean &          1.25 &  0.001 \\
       &                 Median &          1.12 &       0.001 \\
       &     Standard deviation &          0.62 &      0.001 \\
       &    Percentage positive &        100.00 &         - \\
       & Percentage significant &        100.00 &         - \\

\midrule
\multirow{5}{*}{5} &                   Mean &          0.86 &     0.001 \\
       &                 Median &          0.77 &     0.001 \\
       &     Standard deviation &          0.42 &     0.000 \\
       &    Percentage positive &        100.00 &       - \\
       & Percentage significant &        100.00 &       - \\

\midrule
\multirow{5}{*}{10} &                   Mean &          0.46 &     0.001 \\
      &                 Median &          0.41 &     0.001 \\
      &     Standard deviation &          0.22 &     0.000 \\
      &    Percentage positive &        100.00 &       - \\
      & Percentage significant &        100.00 &       - \\

\bottomrule
\end{tabularx}
    \label{tab:mrqap_residual}
\end{table}
}


\section{High-dimensional covariance estimation and application to portfolio allocation} \label{sec:covariance_estimation}
 
The realized covariance matrices 
in Section \ref{sec:co-trading_covariance} pertain to 457 stocks, but are constructed from 78 samples (corresponding to 5 minute buckets) for each stock. The \gdr{parameter} estimations \gdr{thus} fall into a high-dimensional setting where the number of samples is smaller than the dimension of the covariance matrix, resulting in singular estimates. However, it is important to have invertible and well-behaved estimates in practice, \mc{for tasks such as} mean-variance portfolio allocation. 

Fortunately, we have 
\gr{found} significant associations between realized covariance, \update{especially its idiosyncratic component}, and co-trading matrices. With the aid of data-driven co-trading clusters, we develop a method for robust estimation of high-dimensional covariance matrices. This approach is motivated \mc{by the work of  
\cite{ait2017using}, and can be construed as an extension of it that considers higher-order interactions by leveraging very granular high-frequency data \update{\mcc{that incorporates the} dynamic topological structures of \mcc{the network of} stocks}.}

\subsection{Robust covariance matrix estimation}


\update{With the proven capability of capturing dynamic communities in the market and patterns in idiosyncratic covariance}, we propose a robust covariance estimator achieving positive definiteness by imposing a block structure on the idiosyncratic part of the sample covariance matrix, $\Hat{\mathbf{\Sigma}}^{u}$, based on the data-driven clusters derived from co-trading matrices. \update{We first estimate the covariance matrix as in \Cref{sec:co-trading_covariance}, and decompose each realized covariance matrix into factor and idiosyncratic components, as in \eqref{eq:decompose_covariance}. Next, we} threshold the idiosyncratic covariance matrix to obtain a sparse matrix $\Hat{\mathbf{\Gamma}}^{u}$, whose element corresponding to the pair of stocks $i$ and $j$ is
\begin{equation*}
    \Hat{\mathbf{\Gamma}}^{u}_{i,j} = 
    \begin{cases}
        \Hat{\mathbf{\Sigma}}^{u}_{i, j}, & \text{if stock $i$ and stock $j$ belong to the same cluster}  \\
        0, & \text{otherwise}.
    \end{cases}
\end{equation*}
Our \gr{proposed covariance} estimator, incorporating the co-trading clusters, is then given by
\begin{equation*}
    \mathbf{\Hat{\Sigma}}^{Cluster} = \sum_{k=1}^{K} \Hat{\lambda}_{k}^{} \Hat{\mathbf{v}}_{k}^{} \Hat{\mathbf{v}}_{k}^{T} + \Hat{\mathbf{\Gamma}}^{u}. 
\end{equation*}
The estimator is positive definite if the size of every cluster is smaller than the rank of the singular sample covariance matrix, which is usually the number of samples used to calculate it. To be specific, 
with a sampling frequency of 5 minutes, the largest cluster should contain no more than 78 stocks. 
\gr{In contrast to the paper} \cite{ait2017using}, which uses GICS sectors as a fixed ``cluster", we can tune the number of clusters to guarantee positive definiteness given any universe of stocks. 
\gr{Moreover}, thresholding with co-trading clusters embeds \gr{the} dynamic of the dependency structure in stock markets, which is 
reasonable for covariance estimation at daily or higher frequency.  

\subsection{Mean-variance portfolio construction} 
To demonstrate economic value of the proposed robust covariance estimates, we develop a daily rebalanced mean-variance strategy, which opens positions at market open and liquidates at market close, for each trading day over the period of study. Based upon the assumption 
that $\mathbb{E} [\mathbf{\Hat{\Sigma}}_{t} | \mathbf{\Hat{\Sigma}}_{t-1}] = \mathbf{\Hat{\Sigma}}_{t-1} \in \mathbb{R}^{N \times N}$, we derive mean-variance portfolio weights $\mathbf{w}_{t} \in \mathbb{R}^{N}$, on day $t$, by solving the following constrained optimization

\begin{equation} \label{eq:mv_opt}
    \begin{aligned}
        \min_{w_t} \quad & \mathbf{w}_{t}^{T} \mathbf{\Hat{\Sigma}}_{t-1} \mathbf{w}_{t} \\
        \textit{s.t. } & \mathbf{w}_{t}^{T} \vec{\mathbf{1}} = 1 \\
                     & ||\mathbf{w}_{t}||_{1} \leq \textit{l} ,  \\
    \end{aligned}
\end{equation}
\mc{where $\vec{\mathbf{1}}$ denotes the all-ones vector}, and  $l \geq 0$ restricts the level of leverage. When $l = 1$, leverage is not allowed and all weights are non-negative. In contrast, when $l = \infty$, short-sell is unrestricted and the optimization leads to the global minimum variance portfolio (GMV) (\mc{\cite{jagannathan2003risk, bollerslev2018modeling}}). 
%
In the special case that $l = \infty$, 
an analytical solution, $w_{t} = \frac{\mathbf{\Hat{\Sigma}}_{t-1}^{-1} \vec{\mathbf{1}}}{\vec{\mathbf{1}}^{T} \mathbf{\Hat{\Sigma}}_{t-1}^{-1} \vec{\mathbf{1}}}$, to the constrained optimization problem exists. Otherwise, we solve it numerically using the `CVXPY' package in Python (\cite{diamond2016cvxpy, agrawal2018rewriting}). Note that it is possible for $\mathbf{\Hat{\Sigma}}_{t-1}$ to be numerically singular. To avoid this issue, we do not trade on days with ill-behaved covariance estimates, and  
\gr{we set} $\mathbf{w}_{t} = \vec{\mathbf{0}}$ if the condition number of $\mathbf{\Hat{\Sigma}}_{t-1}$ is greater than $1 \times 10^{9}$. Finally, the daily portfolio return is calculated as 
\begin{equation*}
    r_{t}^{mv} = \mathbf{w}_{t}^{T} \mathbf{r}_{t},
\end{equation*}
where $\mathbf{r}_{t} \in \mathbb{R}^{N}$ is a vector of stock logarithmic open-to-close returns.

The evaluation metrics of portfolio performance are annualized volatility and Sharpe ratio. The annualized volatility is calculated as 
\begin{equation*}
    \sigma^{mv} = \textit{stdev}(r_{t}^{mv}) \times \sqrt{252},
\end{equation*}
where $\textit{stdev}(\cdot)$ is the standard deviation function. The annualized Sharpe ratio (\cite{sharpe1994sharpe}) is defined as
\begin{equation*}
    sr^{mv} = \frac{\overline{{r}^{mv}} \times 252}{\sigma^{mv}},
\end{equation*}
where $\overline{{r}^{mv}}$ denotes the average daily return of the portfolio. 
\gdr{A} high Sharpe ratio indicates lower portfolio volatility, adjusted by mean return. 

\subsection{Analysis of portfolio performance}
In the following empirical portfolio analysis, \mc{we experiment with multiple values of the parameters corresponding to the} number of data-driven clusters, number of factors for covariance estimation, and leverage constraints for optimization. For each set of parameters, we construct a daily  portfolio, and perform a backtest over the period of study. Since we calculate portfolio weights traded on day $t$ from  covariance estimates from day $t-1$, all of our tests are out-of-sample. 

\begin{figure}[htp]
    \centering
    \includegraphics[width = \textwidth]{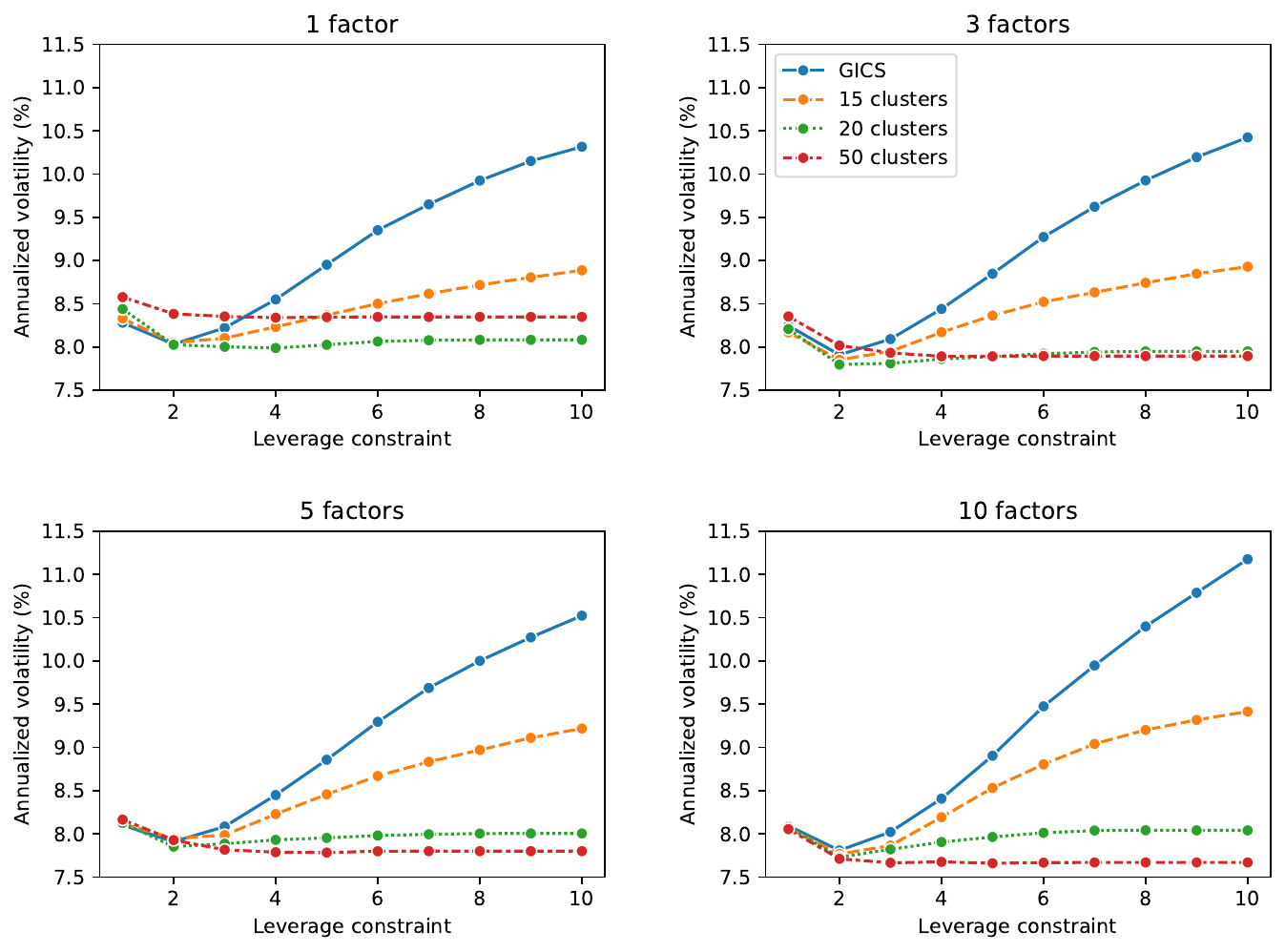}
    \vspace{-4mm}  
    \caption{Annualized volatility of mean-variance portfolios. \\ These figures show the annualized volatility of mean-variance portfolios constructed by solving \eqref{eq:mv_opt} based on different covariance matrices estimates. The out-of-sample backtests span the period from 2017-01-03 to 2019-12-09. Every sub-figure corresponds to one choice of latent factor numbers while decomposing sample covariance matrices in \eqref{eq:decompose_covariance}.
    In each sub-figure, every curve plots portfolio volatility, for a choice of blocks structure imposed on the residual covariance, along leverage constraints.} 
    \label{fig:mv_volatility}
\end{figure}

The annualized volatility of the portfolios is shown in Figure \ref{fig:mv_volatility}. We conclude that a larger number of clusters, which imposes higher level of sparsity on the residual covariance components, leads to more robust covariance estimation, and lower variation in portfolio returns. With short-sell prohibited, the portfolio risks are similar. As leverage constraint\gr{s} relax, the GICS portfolios with 11 fixed clusters and the portfolios corresponding to 15 co-trading clusters \mc{exhibit increasing annualized volatility}.
In contrast, when we consider 20 and 50 co-trading clusters, the annualized portfolio risks decrease, and stabilize at around $8\%$, regardless of the number of factors. 

Figure \ref{fig:mv_volatility} thus indicates that portfolios using our method of robust covariance estimation can lead to superior performance compared to portfolios built upon GICS sector-based covariance estimates.
For further comparison after adjusting for returns, we add Sharpe ratios for selected portfolios in Table \ref{tab:mv_sharpe}. \gr{In alignment with} 
with our findings \gr{from  Figure \ref{fig:mv_volatility}}, data-driven clusters outperform fixed GICS sectors for robust covariance matrix estimation. However, \mc{increasing} the number of clusters too much deteriorates \gr{the} Sharpe ratio. It is noteworthy that, with 50 clusters, the idiosyncratic covariance matrices are overly sparse and the covariance among stocks is underestimated. Thus the portfolio optimization tends to select stocks with lower volatility instead of diversifying, and results in low level of returns. 
\mc{From  the numbers of clusters considered here, the highest Sharpe ratio of 1.40 is attained by the GMV for 20 clusters with 10 latent factors.}

\begin{table}[htp]
    \caption{Annualized Sharpe ratio of mean-variance portfolios. \\ This table documents the annualized Sharpe ratios of mean-variance portfolios constructed by solving \eqref{eq:mv_opt} based on different covariance matrices estimates. The out-of-sample backtests span the period from 2017-01-03 to 2019-12-09. \gr{The} `Factor' column specifies the number of latent factors while decomposing the sample covariance matrices.  The `Leverage' column indicates leverage constraints in \eqref{eq:mv_opt}, where $\infty$ means that short-sell is unrestricted. We use 15, 20 and 50 clusters, together with GICS sectors as the baseline, while imposing \mc{diagonal block structure} on the residual covariance matrices.} 
    \begin{tabularx}{\textwidth}{l*{6}{Y}}
\toprule
Factor              & Leverage & \multicolumn{4}{c}{Cluster}   \\
\midrule
                    &           & GICS     & 15   & 20   & 50    \\
 \cmidrule(lr){3-6}
\multirow{5}{*}{1}  & 1        & 0.01 & 0.13 & 0.09 & -0.06   \\
                    & 3        & 0.57 & 0.61 & 0.53 & 0.12    \\
                    & 5        & 0.75 & 0.89 & 0.69 & 0.12    \\
                    & 7        & 0.71 & 0.98 & 0.69 & 0.12    \\
                    & $\infty$ & 0.45 & \bftab 1.07 & 0.69 & 0.12   \\
\midrule
\multirow{5}{*}{3}  & 1        & 0.35 & 0.48 & 0.44 & 0.36    \\
                    & 3        & 0.54 & 0.74 & 0.86 & 0.51    \\
                    & 5        & 0.95 & 1.02 & 1.04 & 0.54    \\
                    & 7        & 0.92 & 1.13 & 1.11 & 0.54    \\
                    & $\infty$ & 0.59 & \bftab 1.19 & 1.12 & 0.54    \\
\midrule
\multirow{5}{*}{5}  & 1        & 0.38 & 0.52 & 0.51 & 0.47    \\
                    & 3        & 0.63 & 0.81 & 0.88 & 0.62    \\
                    & 5        & 1.04 & 1.11 & 1.16 & 0.70     \\
                    & 7        & 1.06 & 1.28 & 1.23 & 0.71    \\
                    & $\infty$ & 0.69 & \bftab 1.29 & 1.23 & 0.71    \\
\midrule
\multirow{5}{*}{10} & 1        & 0.44 & 0.50 & 0.45 & 0.42    \\
                    & 3        & 0.64 & 0.89 & 1.04 & 0.58    \\
                    & 5        & 1.12 & 1.16 & 1.34 & 0.69    \\
                    & 7        & 1.27 & 1.28 & 1.39 & 0.70     \\
                    & $\infty$ & 0.97 & 1.18 & \bftab 1.40  & 0.70  \\
\bottomrule
\end{tabularx}

    \label{tab:mv_sharpe}
\end{table}

\begin{figure}[htp]
    \centering
    \includegraphics[width = \textwidth]{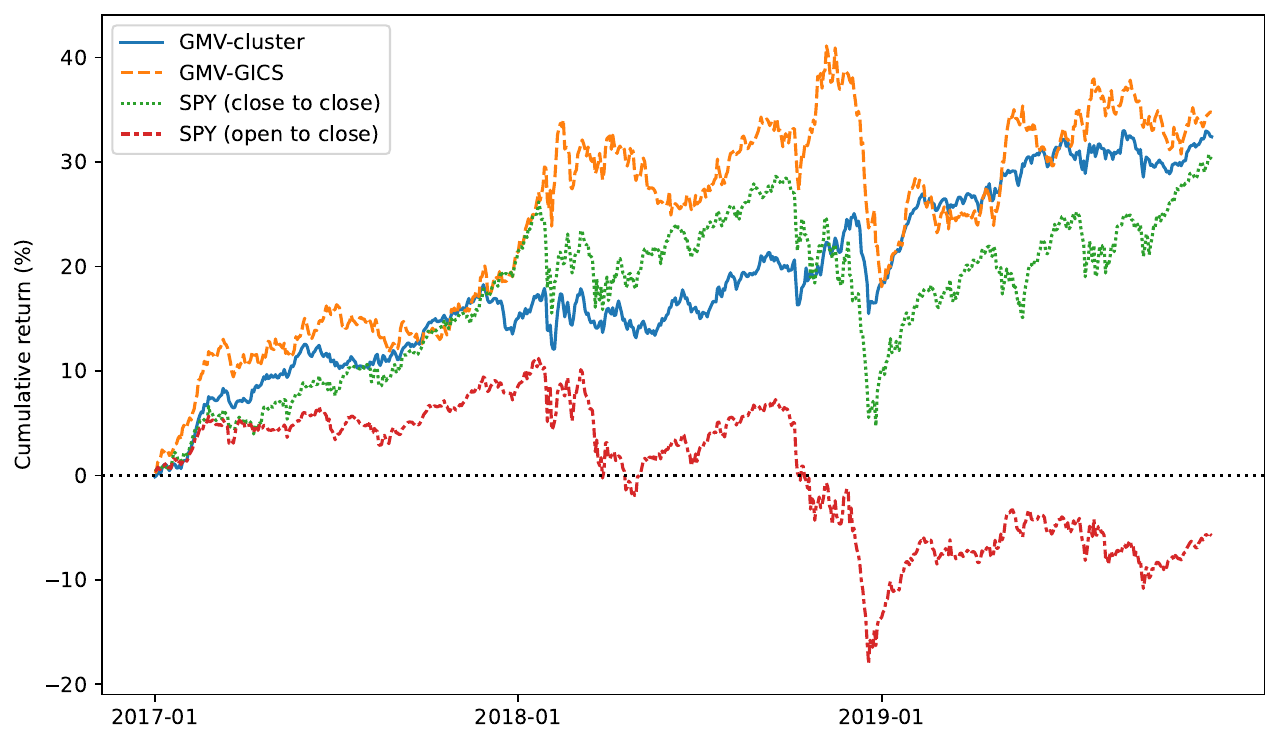}
    \vspace{-4mm}  
    \caption{Cumulative returns of portfolios. \\ This figure plots cumulative returns of four portfolios from 2017-01-03 to 2019-12-09. The portfolios include (1) `GMV-cluster': the global minimum variance portfolio based on robust covariance matrices estimated using our method with 10 factors and 20 clusters; (2) `GMV-GICS': the GMV portfolio based on robust covariance matrices estimated using 10 factors and GICS sectors; (3) `SPY (close to close)': the SPDR S$\&$P 500 ETF Trust which tracks the S$\&$P 500 Index; (4) `SPY (open to close)': the SPY ETF, following our strategy, with positions only during normal trading hours.} 
    \label{fig:mv_benchmark_pnl}
\end{figure}

Zooming into the trajectory of portfolio gains, we sketch the cumulative returns of the best GMV portfolios for the data-driven clusters we propose, along with the GICS sector as a baseline, respectively, in Figure \ref{fig:mv_benchmark_pnl}. Furthermore, we add the \mc{S\&P 500 index (SPY ETF)} 
as a proxy for the market. Clearly, trading SPY from open to close every day \mc{is not a profitable trading strategy}. Apart from that, the portfolio derived from our method outperforms the GICS baseline and the market benchmark, by achieving comparable profits while bearing \mc{much smaller fluctuations and shorter drawdown periods}.

\section{Robustness analysis} \label{sec:robustness}
In this section, we briefly discuss the robustness of \gr{the} spectral clustering algorithm. 
\gr{Moreover, we 
report on the results when}  \mc{calculating co-trading scores based on traded volume instead of the number of trades}. Further details of robustness checks are provided in the Appendix. 

\subsection{Random initialization of spectral clustering algorithm}
\mc{The spectral clustering algorithm applies K-means (\cite{macqueen1967classification}) on the spectral domain of the co-trading matrices. 
\gdr{Thus}, one first computes a low-dimensional embedding of the stocks given by the extremal eigenvectors of the graph Laplacian matrix associated to the co-trading network, and then performs clustering via K-means (\cite{von2007tutorial}).
An issue of K-means is that the clustering results may be sensitive to \gdr{the} random initialization. In order to mitigate this sensitivity, we adopt K-means++ (\cite{arthur2006k}) for \gdr{the} initialization. 
To further empirically examine the robustness, we repeat the experiment of clustering on daily co-trading networks in \Cref{sec:daily_clustering} in 100 trials, each time with a different random seed. Then we compare the daily means of ARIs between clusters from each pair of experiments. Our results ensure that the spectral clustering method we use is robust to finding clusters in co-trading networks. Further details are provided} in \Cref{appendix_clustering_initialization}.

\subsection{Volume-driven co-trading behavior}

Instead of 
incorporating the number of transactions, we also explore the possibility of defining co-trading in terms of traded volumes. Details 
\gdr{can be found} in \Cref{appendix_volume}. By comparing the ARIs between corresponding clusters and GICS sectors, we observe that volume based co-trading matrices share the same patterns as those measured in the number of transactions. 
According to the portfolio performance, described in \Cref{sec:covariance_estimation}, the Sharpe ratios of GMV portfolios corresponding to the volume-based co-trading matrices surpass the GICS benchmarks. \gr{Overall, o}ur findings are robust under the volume measure,  \gr{in the sense that they show similar patterns}. 

However, compared with the count of trades, using \gdr{traded} volume in measuring co-trading is less robust in terms of daily random seed tests and dynamic comparison with GICS sectors. Moreover, volume based co-trading matrices lead to inferior economic value compared to count-based matrices, when using the mean-variance portfolio analysis in \Cref{sec:covariance_estimation}. These findings echo previous research (\cite{chan1995behavior, chordia2004order}) which shows that, since institutions tend to split large orders to hide 
their liquidation purpose and reduce market impact, the number of transactions outperforms the volume in measuring price impact. We confirm that using count of trades better captures patterns in price co-movement.

\section{Conclusion and future research directions} \label{sec:conclusion}
In this paper, we
\gr{introduce and construct} co-trading networks to model the dependency structure of stocks arising from the interaction \gr{of cross-stock trading} among market participants, 
in response of trade arrivals on the market. Using a spectral clustering algorithm, we \mc{uncover} 
clusters which capture \mc{well temporally evolving structures within  markets}, containing information beyond industry sectors. Our empirical studies, focusing on daily co-trading \gr{during  2017-01-03 to 2019-12-09}, reveal that cross-stock trading behaviors are time-varying, and co-trading relations across different GICS sectors become apparent. These 
\gr{two observations} indicate that solely relying on sectors is not sufficient \gr{for capturing co-trading behavior}, and \gr{they motivate the construction and analysis of}
time series of co-trading networks \mc{built from very granular high-frequency data}.

Taking 
\gdr{one step} further,
we \mc{establish} 
that strong co-trading relations \gr{can} lead to a high level of co-movements in stock prices. We 
\gr{use} realized covariance matrices \gr{to} measure the 
co-movements \gr{of prices}. Through network regression analysis, we document a significant positive relation between co-trading and covariance matrices, \update{especially idiosyncratic covariance matrices}. Even \gr{when} adding a network of sectors as a control variable in \gr{the} regression, the conclusion remains valid. This conclusion bridges the gap between cross-stock trading activities at the microstructure level and \gr{the} macroscopic covariance of stock returns.

Employing dynamic co-trading networks and data-driven clusters, we develop a robust co-variance estimator for stock covariance, in a situation where \gr{the number of} samples for estimating \gr{the} covariance is smaller than the number of stocks. Our method outputs well-behaved estimates from sample covariance matrices, by incorporating contemporaneous information of stock clusters. As a result, \gr{a} mean-variance portfolio constructed with our robust estimates achieve\gr{s} lower volatility and higher Sharpe ratios in comparison with baseline methods and market returns.

\gr{O}ur concept of co-trading \mc{provides a 
general framework}  \gr{to}    investigate \mc{the interaction of trade flow corresponding to different stocks; for example, one could take into account the directions of trades, as defined in \Cref{sec:co-trading_network}, when analyzing the co-trading behavior.}
It 
\gr{would also be} worthwhile to further investigate how trades with same (resp., opposite) directions contribute to the positive (resp.,  negative) components of covariance and correlation of stocks. Additionally, in this study we assume that  the co-trading score is symmetric; however, 
\gr{asymmetric scores may also be of interest}, \mc{and the pairwise relationships could be modeled and clustered using methodology from the directed graph clustering literature (\cite{cucuringu2020hermitian})}.  
\gdr{For convenience this paper assumes that the universe of stocks does not change across time; it could be interesting to include the possibility of some new stocks arriving and some existing stocks ceasing to trade.} 
Intuitively, shocks on companies with large \mc{market cap} \gr{can} have impact on small cap stocks, but the converse is \gr{often} not true. With a simple adjustment to the current co-trading scores, this 
\mc{asymmetry}  
\gr{could} be embedded into the network \mc{construction}, \mc{allowing for the investigation of 
asymmetric spillover effects}. \mc{Furthermore, 
it would be interesting to leverage the co-trading network time series for}
forecasting tasks concerning market structures, returns, \gr{and} covariances, 
\gr{possibly} combined with models ranging from parsimonious to deep learning.

\printbibliography

@article{capponi2020multi,
  title={Multi-asset market impact and order flow commonality},
  author={Capponi, Francesco and Cont, Rama},
  journal={Available at SSRN},
  year={2020}
}

@article{chordia2004order,
  title={Order imbalance and individual stock returns: Theory and evidence},
  author={Chordia, Tarun and Subrahmanyam, Avanidhar},
  journal={Journal of Financial Economics},
  volume={72},
  number={3},
  pages={485--518},
  year={2004},
  publisher={Elsevier}
}

@article{hubert1985comparing,
  title={Comparing partitions},
  author={Hubert, Lawrence and Arabie, Phipps},
  journal={Journal of Classification},
  volume={2},
  number={1},
  pages={193--218},
  year={1985},
  publisher={Springer}
}

@article{huang2011lobster,
  title={Lobster: Limit order book reconstruction system},
  author={Huang, Ruihong and Polak, Tomas},
  journal={Available at SSRN 1977207},
  year={2011}
}

@article{ross1976arbitrage,
  title={The arbitrage theory of capital asset pricing},
  author={Ross, Stephen A},
  journal={Journal of Economic Theory},
  volume={13},
  number={3},
  pages={341--360},
  year={1976},
  publisher={Elsevier}
}

@article{sharpe1964capital,
  title={Capital asset prices: A theory of market equilibrium under conditions of risk},
  author={Sharpe, William F},
  journal={The Journal of Finance},
  volume={19},
  number={3},
  pages={425--442},
  year={1964},
  publisher={Wiley Online Library}
}

@article{fama1993common,
  title={Common risk factors in the returns on stocks and bonds},
  author={Fama, Eugene F and French, Kenneth R},
  journal={Journal of Financial Economics},
  volume={33},
  number={1},
  pages={3--56},
  year={1993},
  publisher={Elsevier}
}

@article{kyle1985continuous,
  title={Continuous auctions and insider trading},
  author={Kyle, Albert S},
  journal={Econometrica: Journal of the Econometric Society},
  pages={1315--1335},
  year={1985},
  publisher={JSTOR}
}

@article{hirschey2021high,
  title={Do high-frequency traders anticipate buying and selling pressure?},
  author={Hirschey, Nicholas},
  journal={Management Science},
  volume={67},
  number={6},
  pages={3321--3345},
  year={2021},
  publisher={INFORMS}
}

@article{mantegna1999hierarchical,
  title={Hierarchical structure in financial markets},
  author={Mantegna, Rosario N},
  journal={The European Physical Journal B-Condensed Matter and Complex Systems},
  volume={11},
  number={1},
  pages={193--197},
  year={1999},
  publisher={Springer}
}

@article{von2007tutorial,
  title={A tutorial on spectral clustering},
  author={Von Luxburg, Ulrike},
  journal={Statistics and Computing},
  volume={17},
  number={4},
  pages={395--416},
  year={2007},
  publisher={Springer}
}

@article{shi2000normalized,
  title={Normalized cuts and image segmentation},
  author={Shi, Jianbo and Malik, Jitendra},
  journal={IEEE Transactions on Pattern Analysis and Machine Intelligence},
  volume={22},
  number={8},
  pages={888--905},
  year={2000},
  publisher={Ieee}
}

@inproceedings{ng2002spectral,
  title={On spectral clustering: Analysis and an algorithm},
  author={Ng, Andrew Y and Jordan, Michael I and Weiss, Yair},
  booktitle={Advances in Neural Information Processing Systems},
  pages={849--856},
  year={2002}
}

@inproceedings{cucuringu2020hermitian,
  title={Hermitian matrices for clustering directed graphs: insights and applications},
  author={Cucuringu, Mihai and Li, Huan and Sun, He and Zanetti, Luca},
  booktitle={International Conference on Artificial Intelligence and Statistics},
  pages={983--992},
  year={2020},
  organization={PMLR}
}

@inproceedings{cucuringu2019sponge,
  title={SPONGE: A generalized eigenproblem for clustering signed networks},
  author={Cucuringu, Mihai and Davies, Peter and Glielmo, Aldo and Tyagi, Hemant},
  booktitle={The 22nd International Conference on Artificial Intelligence and Statistics},
  pages={1088--1098},
  year={2019},
  organization={PMLR}
}

@article{fama1992cross,
  title={The Cross-Section of Expected Stock Returns},
  author={Fama, Eugene F and French, Kenneth R},
  journal={The Journal of Finance},
  volume={47},
  number={2},
  pages={427--465},
  year={1992}
}

@article{farine2013animal,
  title={Animal social network inference and permutations for ecologists in R using asnipe},
  author={Farine, Damien R},
  journal={Methods in Ecology and Evolution},
  volume={4},
  number={12},
  pages={1187--1194},
  year={2013},
  publisher={Wiley Online Library}
}

@article{sharpe1994sharpe,
  title={The sharpe ratio},
  author={Sharpe, William F},
  journal={Journal of Portfolio Management},
  volume={21},
  number={1},
  pages={49--58},
  year={1994},
  publisher={INSTITUTIONAL INVESTOR INC 488 MADISON AVENUE, NEW YORK, NY 10022}
}

@article{huang2009network,
  title={A network analysis of the Chinese stock market},
  author={Huang, Wei-Qiang and Zhuang, Xin-Tian and Yao, Shuang},
  journal={Physica A: Statistical Mechanics and its Applications},
  volume={388},
  number={14},
  pages={2956--2964},
  year={2009},
  publisher={Elsevier}
}

@article{namaki2011network,
  title={Network analysis of a financial market based on genuine correlation and threshold method},
  author={Namaki, Ali and Shirazi, Amir H and Raei, R and Jafari, GR},
  journal={Physica A: Statistical Mechanics and its Applications},
  volume={390},
  number={21-22},
  pages={3835--3841},
  year={2011},
  publisher={Elsevier}
}

@article{tumminello2005tool,
  title={A tool for filtering information in complex systems},
  author={Tumminello, Michele and Aste, Tomaso and Di Matteo, Tiziana and Mantegna, Rosario N},
  journal={Proceedings of the National Academy of Sciences},
  volume={102},
  number={30},
  pages={10421--10426},
  year={2005},
  publisher={National Acad Sciences}
}

@article{mcdonald2005detecting,
  title={Detecting a currency’s dominance or dependence using foreign exchange network trees},
  author={McDonald, Mark and Suleman, Omer and Williams, Stacy and Howison, Sam and Johnson, Neil F},
  journal={Physical Review E},
  volume={72},
  number={4},
  pages={046106},
  year={2005},
  publisher={APS}
}

@article{nie2017dynamics,
  title={Dynamics of cluster structure in financial correlation matrix},
  author={Nie, Chun-Xiao},
  journal={Chaos, Solitons \& Fractals},
  volume={104},
  pages={835--840},
  year={2017},
  publisher={Elsevier}
}

@article{van2019high,
  title={High-frequency trading around large institutional orders},
  author={Van Kervel, Vincent and Menkveld, Albert J},
  journal={The Journal of Finance},
  volume={74},
  number={3},
  pages={1091--1137},
  year={2019},
  publisher={Wiley Online Library}
}

@article{fama2015five,
  title={A five-factor asset pricing model},
  author={Fama, Eugene F and French, Kenneth R},
  journal={Journal of Financial Economics},
  volume={116},
  number={1},
  pages={1--22},
  year={2015},
  publisher={Elsevier}
}

@article{chan1995behavior,
  title={The behavior of stock prices around institutional trades},
  author={Chan, Louis KC and Lakonishok, Josef},
  journal={The Journal of Finance},
  volume={50},
  number={4},
  pages={1147--1174},
  year={1995},
  publisher={Wiley Online Library}
}

@article{grossman1988liquidity,
  title={Liquidity and market structure},
  author={Grossman, Sanford J and Miller, Merton H},
  journal={The Journal of Finance},
  volume={43},
  number={3},
  pages={617--633},
  year={1988},
  publisher={Wiley Online Library}
}

@article{brunnermeier2005predatory,
  title={Predatory trading},
  author={Brunnermeier, Markus K and Pedersen, Lasse Heje},
  journal={The Journal of Finance},
  volume={60},
  number={4},
  pages={1825--1863},
  year={2005},
  publisher={Wiley Online Library}
}

@article{yang2020back,
  title={Back-running: Seeking and hiding fundamental information in order flows},
  author={Yang, Liyan and Zhu, Haoxiang},
  journal={The Review of Financial Studies},
  volume={33},
  number={4},
  pages={1484--1533},
  year={2020},
  publisher={Oxford University Press}
}

@article{fan2016incorporating,
  title={Incorporating global industrial classification standard into portfolio allocation: A simple factor-based large covariance matrix estimator with high-frequency data},
  author={Fan, Jianqing and Furger, Alex and Xiu, Dacheng},
  journal={Journal of Business \& Economic Statistics},
  volume={34},
  number={4},
  pages={489--503},
  year={2016},
  publisher={Taylor \& Francis}
}

@article{fan2016overview,
  title={An overview of the estimation of large covariance and precision matrices},
  author={Fan, Jianqing and Liao, Yuan and Liu, Han},
  journal={The Econometrics Journal},
  volume={19},
  number={1},
  pages={C1--C32},
  year={2016},
  publisher={Oxford University Press Oxford, UK}
}

@article{krackhardt1988predicting,
  title={Predicting with networks: Nonparametric multiple regression analysis of dyadic data},
  author={Krackhardt, David},
  journal={Social Networks},
  volume={10},
  number={4},
  pages={359--381},
  year={1988},
  publisher={Elsevier}
}

@article{dekker2007sensitivity,
  title={Sensitivity of MRQAP tests to collinearity and autocorrelation conditions},
  author={Dekker, David and Krackhardt, David and Snijders, Tom AB},
  journal={Psychometrika},
  volume={72},
  number={4},
  pages={563--581},
  year={2007},
  publisher={Springer}
}

@article{mantel1967detection,
  title={The detection of disease clustering and a generalized regression approach},
  author={Mantel, Nathan},
  journal={Cancer Research},
  volume={27},
  number={2\_Part\_1},
  pages={209--220},
  year={1967},
  publisher={AACR}
}

@article{ait2017using,
  title={Using principal component analysis to estimate a high dimensional factor model with high-frequency data},
  author={Ait-Sahalia, Yacine and Xiu, Dacheng},
  journal={Journal of Econometrics},
  volume={201},
  number={2},
  pages={384--399},
  year={2017},
  publisher={Elsevier}
}

@article{bollerslev2018modeling,
  title={Modeling and forecasting (un) reliable realized covariances for more reliable financial decisions},
  author={Bollerslev, Tim and Patton, Andrew J and Quaedvlieg, Rogier},
  journal={Journal of Econometrics},
  volume={207},
  number={1},
  pages={71--91},
  year={2018},
  publisher={Elsevier}
}

@article{liu2015does,
  title={Does anything beat 5-minute RV? A comparison of realized measures across multiple asset classes},
  author={Liu, Lily Y and Patton, Andrew J and Sheppard, Kevin},
  journal={Journal of Econometrics},
  volume={187},
  number={1},
  pages={293--311},
  year={2015},
  publisher={Elsevier}
}

@article{lu2022trade,
  title={Trade co-occurrence, trade flow decomposition, and conditional order imbalance in equity markets},
  author={Lu, Yutong and Reinert, Gesine and Cucuringu, Mihai},
  journal={arXiv preprint arXiv:2209.10334},
  year={2022}
}

@article{andersen2001distribution,
  title={The distribution of realized stock return volatility},
  author={Andersen, Torben G and Bollerslev, Tim and Diebold, Francis X and Ebens, Heiko},
  journal={Journal of Financial Economics},
  volume={61},
  number={1},
  pages={43--76},
  year={2001},
  publisher={Elsevier}
}

@article{bennett2022lead,
  title={Lead--lag detection and network clustering for multivariate time series with an application to the US equity market},
  author={Bennett, Stefanos and Cucuringu, Mihai and Reinert, Gesine},
  journal={Machine Learning},
  volume={111},
  number={12},
  pages={4497--4538},
  year={2022},
  publisher={Springer}
}

@article{marti2021review,
  title={A review of two decades of correlations, hierarchies, networks and clustering in financial markets},
  author={Marti, Gautier and Nielsen, Frank and Bi{\'n}kowski, Miko{\l}aj and Donnat, Philippe},
  journal={Progress in Information Geometry},
  pages={245--274},
  year={2021},
  publisher={Springer}
}

@article{kruskal1956shortest,
  title={On the shortest spanning subtree of a graph and the traveling salesman problem},
  author={Kruskal, Joseph B},
  journal={Proceedings of the American Mathematical Society},
  volume={7},
  number={1},
  pages={48--50},
  year={1956},
  publisher={JSTOR}
}

@article{bonacich1972factoring,
  title={Factoring and weighting approaches to status scores and clique identification},
  author={Bonacich, Phillip},
  journal={Journal of Mathematical Sociology},
  volume={2},
  number={1},
  pages={113--120},
  year={1972},
  publisher={Taylor \& Francis}
}

@article{bonacich1987power,
  title={Power and centrality: A family of measures},
  author={Bonacich, Phillip},
  journal={American Journal of Sociology},
  volume={92},
  number={5},
  pages={1170--1182},
  year={1987},
  publisher={University of Chicago Press}
}

@article{clark2013exploratory,
  title={Exploratory trading},
  author={Clark-Joseph, Adam},
  journal={Unpublished job market paper. Harvard University, Cambridge, MA},
  year={2013}
}

@article{bardoscia2021physics,
  title={The physics of financial networks},
  author={Bardoscia, Marco and Barucca, Paolo and Battiston, Stefano and Caccioli, Fabio and Cimini, Giulio and Garlaschelli, Diego and Saracco, Fabio and Squartini, Tiziano and Caldarelli, Guido},
  journal={Nature Reviews Physics},
  pages={1--18},
  year={2021},
  publisher={Nature Publishing Group}
}

@article{krackardt1987qap,
  title={QAP partialling as a test of spuriousness},
  author={Krackardt, David},
  journal={Social networks},
  volume={9},
  number={2},
  pages={171--186},
  year={1987},
  publisher={Elsevier}
}

@techreport{arthur2006k,
  title={k-means++: The advantages of careful seeding},
  author={Arthur, David and Vassilvitskii, Sergei},
  year={2006},
  institution={Stanford}
}

@article{plerou2000random,
  title={A random matrix theory approach to financial cross-correlations},
  author={Plerou, Vasiliki and Gopikrishnan, Parameswaran and Rosenow, Bernd and Amaral, LA Nunes and Stanley, H Eugene},
  journal={Physica A: Statistical Mechanics and its Applications},
  volume={287},
  number={3-4},
  pages={374--382},
  year={2000},
  publisher={Elsevier}
}

@article{kullmann2000identification,
  title={Identification of clusters of companies in stock indices via Potts super-paramagnetic transitions},
  author={Kullmann, L and Kertesz, J and Mantegna, RN},
  journal={Physica A: Statistical Mechanics and its Applications},
  volume={287},
  number={3-4},
  pages={412--419},
  year={2000},
  publisher={Elsevier}
}

@article{billio2012econometric,
  title={Econometric measures of connectedness and systemic risk in the finance and insurance sectors},
  author={Billio, Monica and Getmansky, Mila and Lo, Andrew W and Pelizzon, Loriana},
  journal={Journal of Financial Economics},
  volume={104},
  number={3},
  pages={535--559},
  year={2012},
  publisher={Elsevier}
}

@article{fiedor2014information,
  title={Information-theoretic approach to lead-lag effect on financial markets},
  author={Fiedor, Pawe{\l}},
  journal={The European Physical Journal B},
  volume={87},
  number={8},
  pages={1--9},
  year={2014},
  publisher={Springer}
}

@article{ding2021stock,
  title={Stock co-jump networks},
  author={Ding, Yi and Li, Yingying and Liu, Guoli and Zheng, Xinghua},
  journal={Available at SSRN},
  year={2021}
}

@article{ledoit2003improved,
  title={Improved estimation of the covariance matrix of stock returns with an application to portfolio selection},
  author={Ledoit, Olivier and Wolf, Michael},
  journal={Journal of Empirical Finance},
  volume={10},
  number={5},
  pages={603--621},
  year={2003},
  publisher={Elsevier}
}

@article{10.2307/2975974,
 ISSN = {00221082, 15406261},
 author = {Harry Markowitz},
 journal = {The Journal of Finance},
 number = {1},
 pages = {77--91},
 publisher = {[American Finance Association, Wiley]},
 title = {Portfolio selection},
 volume = {7},
 year = {1952}
}

@article{ledoit2004well,
  title={A well-conditioned estimator for large-dimensional covariance matrices},
  author={Ledoit, Olivier and Wolf, Michael},
  journal={Journal of Multivariate Analysis},
  volume={88},
  number={2},
  pages={365--411},
  year={2004},
  publisher={Elsevier}
}

@article{bickel2008regularized,
  title={Regularized estimation of large covariance matrices},
  author={Bickel, Peter J and Levina, Elizaveta},
  journal={The Annals of Statistics},
  volume={36},
  number={1},
  pages={199--227},
  year={2008},
  publisher={Institute of Mathematical Statistics}
}

@article{bickel2008covariance,
  title={Covariance regularization by thresholding},
  author={Bickel, Peter J and Levina, Elizaveta},
  journal={The Annals of Statistics},
  volume={36},
  number={6},
  pages={2577--2604},
  year={2008},
  publisher={Institute of Mathematical Statistics}
}

@article{chen2010shrinkage,
  title={Shrinkage algorithms for MMSE covariance estimation},
  author={Chen, Yilun and Wiesel, Ami and Eldar, Yonina C and Hero, Alfred O},
  journal={IEEE Transactions on Signal Processing},
  volume={58},
  number={10},
  pages={5016--5029},
  year={2010},
  publisher={IEEE}
}

@article{jagannathan2003risk,
  title={Risk reduction in large portfolios: Why imposing the wrong constraints helps},
  author={Jagannathan, Ravi and Ma, Tongshu},
  journal={The Journal of Finance},
  volume={58},
  number={4},
  pages={1651--1683},
  year={2003},
  publisher={Wiley Online Library}
}

@article{bernhardt2008cross,
  title={Cross-asset speculation in stock markets},
  author={Bernhardt, Dan and Taub, Bart},
  journal={The Journal of Finance},
  volume={63},
  number={5},
  pages={2385--2427},
  year={2008},
  publisher={Wiley Online Library}
}

@article{hasbrouck2001common,
  title={Common factors in prices, order flows, and liquidity},
  author={Hasbrouck, Joel and Seppi, Duane J},
  journal={Journal of Financial Economics},
  volume={59},
  number={3},
  pages={383--411},
  year={2001},
  publisher={Elsevier}
}

@article{harford2005correlated,
  title={Correlated order flow: Pervasiveness, sources, and pricing effects},
  author={Harford, Jarrad and Kaul, Aditya},
  journal={Journal of Financial and Quantitative Analysis},
  volume={40},
  number={1},
  pages={29--55},
  year={2005},
  publisher={Cambridge University Press}
}

@article{pasquariello2015strategic,
  title={Strategic cross-trading in the US stock market},
  author={Pasquariello, Paolo and Vega, Clara},
  journal={Review of Finance},
  volume={19},
  number={1},
  pages={229--282},
  year={2015},
  publisher={Oxford University Press}
}

@article{benzaquen2017dissecting,
  title={Dissecting cross-impact on stock markets: An empirical analysis},
  author={Benzaquen, Michael and Mastromatteo, Iacopo and Eisler, Zoltan and Bouchaud, Jean-Philippe},
  journal={Journal of Statistical Mechanics: Theory and Experiment},
  volume={2017},
  number={2},
  pages={023406},
  year={2017},
  publisher={IOP Publishing}
}

@article{schneider2019cross,
  title={Cross-impact and no-dynamic-arbitrage},
  author={Schneider, Michael and Lillo, Fabrizio},
  journal={Quantitative Finance},
  volume={19},
  number={1},
  pages={137--154},
  year={2019},
  publisher={Taylor \& Francis}
}

@inproceedings{macqueen1967classification,
  title={Classification and analysis of multivariate observations},
  author={MacQueen, J},
  booktitle={5th Berkeley Symp. Math. Statist. Probability},
  pages={281--297},
  year={1967},
  organization={University of California Los Angeles LA USA}
}

@techreport{hagberg2008exploring,
  title={Exploring network structure, dynamics, and function using NetworkX},
  author={Hagberg, Aric and Swart, Pieter and S Chult, Daniel},
  year={2008},
  institution={Los Alamos National Lab.(LANL), Los Alamos, NM (United States)}
}

@article{diamond2016cvxpy,
  author  = {Steven Diamond and Stephen Boyd},
  title   = {{CVXPY}: {A} {P}ython-embedded modeling language for convex optimization},
  journal = {Journal of Machine Learning Research},
  year    = {2016},
  volume  = {17},
  number  = {83},
  pages   = {1--5},
}

@article{agrawal2018rewriting,
  author  = {Agrawal, Akshay and Verschueren, Robin and Diamond, Steven and Boyd, Stephen},
  title   = {A rewriting system for convex optimization problems},
  journal = {Journal of Control and Decision},
  year    = {2018},
  volume  = {5},
  number  = {1},
  pages   = {42--60},
}

@article{zhang2023company,
  title={Company competition graph},
  author={Zhang, Yanci and Lu, Yutong and Mao, Haitao and Huang, Jiawei and Zhang, Cien and Li, Xinyi and Dai, Rui},
  journal={arXiv preprint arXiv:2304.00323},
  year={2023}
}

@article{hoberg2016text,
  title={Text-based network industries and endogenous product differentiation},
  author={Hoberg, Gerard and Phillips, Gordon},
  journal={Journal of Political Economy},
  volume={124},
  number={5},
  pages={1423--1465},
  year={2016},
  publisher={University of Chicago Press Chicago, IL}
}

@article{cheng2022financial,
  title={Financial time series forecasting with multi-modality graph neural network},
  author={Cheng, Dawei and Yang, Fangzhou and Xiang, Sheng and Liu, Jin},
  journal={Pattern Recognition},
  volume={121},
  pages={108218},
  year={2022},
  publisher={Elsevier}
}

@article{zhang2023robust,
  title={Robust Detection of Lead-Lag Relationships in Lagged Multi-Factor Models},
  author={Zhang, Yichi and Cucuringu, Mihai and Shestopaloff, Alexander Y and Zohren, Stefan},
  journal={arXiv preprint arXiv:2305.06704},
  year={2023}
}

@article{massara2016network,
  title={Network filtering for big data: Triangulated maximally filtered graph},
  author={Massara, Guido Previde and Di Matteo, Tiziana and Aste, Tomaso},
  journal={Journal of complex Networks},
  volume={5},
  number={2},
  pages={161--178},
  year={2016},
  publisher={Oxford University Press}
}

@article{tumminello2007correlation,
  title={Correlation based networks of equity returns sampled at different time horizons},
  author={Tumminello, Michele and Di Matteo, Tiziana and Aste, Tomaso and Mantegna, Rosario N},
  journal={The European Physical Journal B},
  volume={55},
  pages={209--217},
  year={2007},
  publisher={Springer}
}

@article{di2010use,
  title={The use of dynamical networks to detect the hierarchical organization of financial market sectors},
  author={Di Matteo, Tiziana and Pozzi, Francesca and Aste, Tomaso},
  journal={The European Physical Journal B},
  volume={73},
  pages={3--11},
  year={2010},
  publisher={Springer}
}

@article{goldstein2023high,
  title={High-frequency trading strategies},
  author={Goldstein, Michael and Kwan, Amy and Philip, Richard},
  journal={Management Science},
  volume={69},
  number={8},
  pages={4413--4434},
  year={2023},
  publisher={INFORMS}
}

@article{schwenkler2019network,
  title={The network of firms implied by the news},
  author={Schwenkler, Gustavo and Zheng, Hannan},
  journal={Boston University Questrom School of Business Research Paper},
  number={3320859},
  year={2019}
}

\appendix

\section{Spectral Clustering} \label{appendix_spectral_clustering}
In this section, we describe the spectral clustering algorithm used to cluster stocks based on the co-trading matrices in Section \ref{cluster_method}.

\gr{We represent} 
an 
undirected graph $G = (V, E)$, where $V$ = $\{ v_1, v_2, ..., v_\gr{N} \}$ is a collection of \gr{$N$} vertices which are data points and $E$ is a set of edges,
\gr{by its} 
(weighted) adjacency matrix $A \in \mathbb{R}^{N \times N}$. 
The degree matrix of $A$, denoted as $D$, is the diagonal matrix \gr{with entries} 
\begin{equation*}
    D_{ii} = \sum_{j = 1, j \neq i}^{N} A_{ij} .
\end{equation*}
The graph Laplacian $L$ is defined as
\begin{equation*}
    L = D - A .
\end{equation*}
We use the degree matrix to normalize $L$, and the normalized version is 
\begin{equation*}
    L_{sym} = D^{-\frac{1}{2}} L D^{-\frac{1}{2}} ,
\end{equation*}
which contains which contains the information of graph connectivity. Then we perform K-means clustering on the matrix of eigenvectors corresponding to the smallest $K$ eigenvalues of $L_{sym}$. \gr{Here, $K$ is the pre-selected number of clusters.} The procedure \gr{is}
summarized in Algorithm \ref{algo_spectral_clustering}. 

\begin{algorithm}[htp] 
\caption{\small Spectral Clustering}
\label{algo_spectral_clustering}
\hspace*{\algorithmicindent} \textbf{Input:} An  $N \times N$ similarity matrix $A$, number of clusters $K$. \\
\hspace*{\algorithmicindent} \textbf{Output:} Clusters $C_1, C_2, ..., C_K$ 
\begin{algorithmic}[1]
\Procedure{Spectral\_Clustering}{$A, K$} 
\State Compute normalized Laplacian $L_{sym}$
\State Compute the eigenvectors $v_1, v_2, ..., v_K$ corresponding to $K$ smallest eigenvalues of $L_{sym}$ 
\State Construct matrix $Q \in \mathbb{R}^{N \times K}$ with $v_1, v_2, ..., v_K$ as columns
\State Form matrix $\Tilde{Q} \in \mathbb{R}^{N \times K}$ by normalizing row vectors of $Q$ to norm 1
\State Apply K-means clustering, with \gr{K}-means$++$ (\cite{arthur2006k}) for random initialization, to assign rows of $\Tilde{Q}$ to clusters $C_1, C_2, ..., C_K$ 
\EndProcedure
\end{algorithmic}
\end{algorithm}

\section{Random Initialization} \label{appendix_clustering_initialization}
\begin{figure}[htp]
    \centering
    \includegraphics[width = \textwidth]{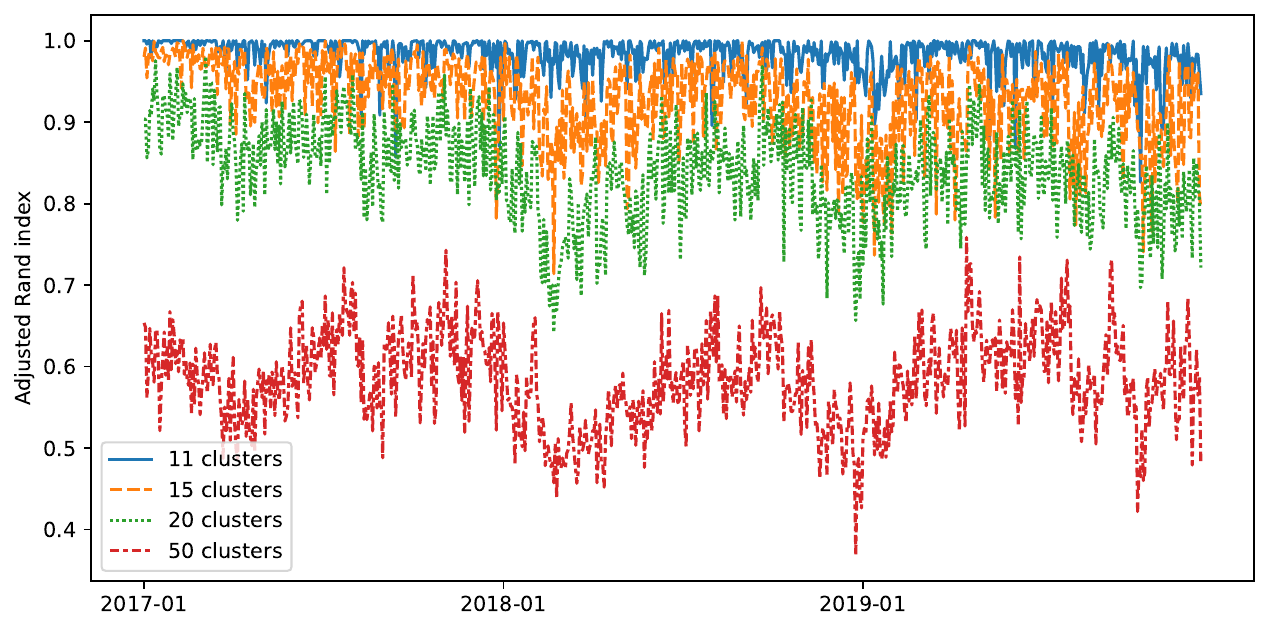}
    \vspace{-4mm}  
    \caption{Robustness of applying spectral clustering on daily co-trading networks. \\ This figure plots the daily means of ARIs between each pair of clusters obtained by running the spectral clustering method 100 times on the same co-trading matrix every day, with different random initialization. Each line corresponds to a value of number of clusters, chosen from 11, 15, 20 and 50.} 
    \label{fig:clustering_robustness}
\end{figure}

Figure \ref{fig:clustering_robustness} plots the average daily ARIs against time. We observe that the values of average daily ARIs are 0.98, 0.93, 0.84 and 0.58, for 11, 15, 20 and 50 clusters respectively. The standard deviations of daily ARIs are 0.11, 0.10, 0.07 and 0.03 over the entire period of study. Hence, our clustering analysis on co-trading matrices is robust to random initialization.

\section{Volume Measure} \label{appendix_volume}
In this section, we define co-trading in terms of volume of trades. As an analogue to \Cref{sec:co-trading_score}, we begin with defining 
\begin{equation*}
    V_{t, j \rightarrow i}^{d^{j} \rightarrow d^{i}} = \sum_{x_k \in S_{t}^{i, d^{i}}} \sum_{x_l \in \{x_a \in \mathcal{N}_{x_k}^{\delta} | s_{a} = j, d_{a} = d^{j\}}} q_l,
\end{equation*}
where $S_{t}^{i, d^{i}}$ is the set of all filtered trades and $q_{l}$ is the size of trade $x_{l}$.

Then, the pairwise volume co-trading score between stock $i$ and stock $j$ on day $t$, using volumes of co-occurred trades for stock $i$ and $j$ with direction $d^j$ and $d^i$, respectively, is defined as 
\begin{equation*}
    c_{t,i,j}^{\delta, d^i, d^j} := \frac{V_{t, i \rightarrow j}^{d^{i} \rightarrow d^{j}} + V_{t, j \rightarrow i}^{d^{j} \rightarrow d^{i}}}{\sqrt{\sum_{x_l \in S_{t}^{i, d^i}} q_l} \sqrt{\sum_{x_m \in S_{t}^{j, d^j}} q_m}}.
\end{equation*}

Similarly, incorporating volumes, a pairwise co-occurrence count index is determined by summing up volumes of co-occurred trades of a pair of stocks and normalizing with \gr{their} total volumes. Finally, we concatenate the pairwise co-trading scores to be the co-trading matrix in volume measure. In line with the \gr{main text of the paper}, 
we set $\delta$ is set to 500 microseconds.

We repeat the empirical analysis and summarize the results in Table \ref{tab:robust_volume}. The portfolios based on volume measure underperform those built upon count measure. Furthermore, the volume measured co-trading matrices are less robust than those measured in count of trades.

\begin{table}[htp]
    \caption{Summary of analysis on volume based co-trading matrices. \\ We construct volume based daily co-trading matrices from 2017-01-03 to 2019-12-09. Panel A reports the annualized Sharpe ratios of global minimum variance (GMV) portfolios constructed by solving \eqref{eq:mv_opt} based on different covariance matrices estimates. \gr{The} `Factor' column specifies the number of latent factors while decomposing the sample covariance matrices.  The `GICS' and `Count-20' columns indicate GMV portfolios corresponding to GICS and clustering count based co-trading matrices with 20 clusters as benchmarks. For volume based co-trading matrices, we use 15, 20 and 50 clusters while imposing \mc{diagonal block structure} on the residual covariance matrices. Additionally, we detect clusters of stocks on volume based co-trading matrices every day with 100 different random seeds for spectral clustering. The predefined number of clusters are 11, 15, 20 and 50. For comparison purpose, we report the statistics for clusters of count based co-trading matrices under. Each value in Panel B is a mean of ARIs between each pair of clusters every day and averaged over all days. Each value in Panel C represents a mean of ARIs between data-driven clusters and GICS sectors each day and averaged over all days.}    
    \begin{tabularx}{\textwidth}{l*{6}{Y}}
\toprule
\multicolumn{6}{l}{\textbf{Panel A: Annualized Sharpe ratios of GMV portfolios.}} \\
\midrule
Factor        &   & \multicolumn{4}{c}{Cluster}   \\
\midrule
              & GICS  & Count-20   & 15   & 20   & 50    \\
 \cmidrule(lr){2-6}
1      & 0.45 &   0.69    & 0.10  & 0.65  & 0.09 \\
3      & 0.59 &   1.12    & 0.36  & 1.00  & 0.63 \\
5      & 0.69 &   1.23    & 0.20  & 0.99  & 0.80 \\
10     & 0.97 &   1.40    & 0.06  & 1.07  & 0.83 \\
\end{tabularx}

\begin{tabularx}{\textwidth}{l*{5}{Y}}
\toprule
\multicolumn{5}{l}{\textbf{Panel B: Average daily ARI among clusters with different initialization.}} \\
\midrule

Measure   & \multicolumn{4}{c}{Cluster}   \\
\midrule
                        & 11     & 15   & 20   & 50    \\
        \cmidrule(lr){2-5} 
Volume                   & 0.97  & 0.90  & 0.80  & 0.52 \\
Count                    & 0.98  & 0.93  & 0.84  & 0.58 \\

\end{tabularx}

\begin{tabularx}{\textwidth}{l*{5}{Y}}
\toprule
\multicolumn{5}{l}{\textbf{Panel C: Average daily ARI between clusters and GICS sectors.}} \\
\midrule

Measure   & \multicolumn{4}{c}{Cluster}   \\
\midrule
                        & 11     & 15   & 20   & 50    \\
        \cmidrule(lr){2-5} 
Volume                   & 0.33  & 0.35  & 0.33  & 0.18 \\
Count                    & 0.41  & 0.43  & 0.40  & 0.21 \\

\bottomrule
\end{tabularx}

    \label{tab:robust_volume}
\end{table}

\end{document}